\documentclass[12pt,letter]{scrartcl}

\usepackage[shortlabels]{enumitem}
\usepackage{amsmath, amssymb, amsthm, amsthm}
\usepackage{xcolor}
\usepackage{graphicx}
\usepackage{adjustbox}
\usepackage{setspace}
\usepackage{cancel}
\usepackage[normalem]{ulem}
\usepackage{units}
\usepackage{microtype}

\usepackage[top=1.0in, right=1.4in, left=1.4in,bottom=1.1in]{geometry}
\usepackage[displaymath, mathlines, pagewise]{lineno}
\usepackage{resizegather}
\usepackage{multirow}
\usepackage{textcomp}
\usepackage{natbib}
\usepackage{mathtools}
\usepackage{bm}
\usepackage{tabularx}
\usepackage{booktabs}
\usepackage{caption}
\usepackage{subcaption}
\usepackage[unicode=true, pdfusetitle,
bookmarks=true,bookmarksnumbered=false,bookmarksopen=false,
breaklinks=false,pdfborder={0 0 1},backref=false,colorlinks=false]
{hyperref}
\hypersetup{
	colorlinks = true,
	urlcolor = blue,
	linkcolor = red,
	citecolor = blue,
}
\usepackage{longtable}
\usepackage{tikz}
\usetikzlibrary{decorations.pathreplacing, arrows.meta}
\tikzset{vert/.style = {circle, fill, inner sep = 0, minimum size = 5}}
\usepackage[multiple]{footmisc}
\usepackage{tabularx}
\usepackage{caption}
\usepackage{subcaption}
\usepackage{multirow}

\newenvironment{tabnotes}[2][1]{\begin{minipage}[t]{#1\textwidth}\vspace{0.1cm}\scriptsize{\emph{Notes:} #2}}{\end{minipage}}

\newtheorem{theorem}{Theorem}
\newtheorem*{theorem*}{Theorem}

\theoremstyle{definition}
\newtheorem{claim}{Claim}
\newtheorem{lemma}{Lemma}

\theoremstyle{definition}
\newtheorem{definition}{Definition}

\newtheorem{corollary}{Corollary}
\theoremstyle{remark}
\newtheorem{example}{Example}

\newcommand{\ver}[1]{{\footnotesize\color{red}#1 \par}}
\renewcommand{\ver}[1]{}
\usepackage{yfonts}
\newcommand{\pti}{{\textfrak{p}}}

\makeatletter
 
 \newcommand*{\defeqi}{=\mathrel{\vcenter{\baselineskip0.5ex \lineskiplimit0pt \hbox{\scriptsize.}\hbox{\scriptsize.}}}}
\makeatother

\usepackage{mathptmx}
\makeatletter
\g@addto@macro\normalsize{%
  \setlength\abovedisplayskip{8pt}
  \setlength\belowdisplayskip{9pt}
  \setlength\abovedisplayshortskip{4pt}
}
\makeatother

\makeatletter
\let\origsection\section
\renewcommand\section{\@ifstar{\starsection}{\nostarsection}}

\newcommand\nostarsection[1]
{\sectionprelude\origsection{#1}\sectionpostlude}

\newcommand\starsection[1]
{\sectionprelude\origsection*{#1}\sectionpostlude}

\newcommand\sectionprelude{%
  \vspace{-1.2em}
}

\newcommand\sectionpostlude{%
  \vspace{-0.2em}
}
\makeatother

\title{Assignment mechanisms: common preferences and information acquisition}
\date{October 17, 2021}
\author{
 Georgy Artemov\thanks{Department of Economics, The University of Melbourne, 111 Barry St., Carlton, VIC 3010, Australia. Tel: +61 (3) 83447029; Email \url{gartemov@unimelb.edu.au}}
}

\begin{document}

\maketitle

\begin{abstract}
I study costly information acquisition in a two-sided matching problem, such as matching applicants to schools. An applicant's utility is a sum of common and idiosyncratic components. The idiosyncratic component is unknown to the applicant but can be learned at a cost. As applicants learn, their preferences over schools become more heterogeneous, improving match quality. In my stylized environment, too few applicants acquire information in an ordinal strategy-proof mechanism. Subsidies, disclosure of applicants' priorities, and affirmative action-like policies lead to higher information acquisition and Pareto improvements. Learning may also decrease when an ordinal strategy-proof mechanism replaces an Immediate Acceptance mechanism.

\textit{JEL\ classification:} D47, D82

\textit{Keywords: School choice, information acquisition, Deferred Acceptance, Immediate Acceptance}
\end{abstract}

\vfill

\noindent
Published article can be found at \url{https://doi.org/10.1016/j.jet.2021.105370}

\noindent
\copyright 2020. This manuscript version is made available under the CC-BY-NC-ND 4.0 license\\ \url{https://creativecommons.org/licenses/by-nc-nd/4.0/}

\newpage
\setlist{nolistsep}

Some applicants -- or their parents -- visit multiple schools, research them online and seek advice from other applicants before settling on the school best suited for them. Others, faced with such a time-consuming endeavor, decide to follow a public ranking of schools, such as compiled by media, without acquiring any applicant-specific information.\footnote{\citet{Hastings/VanWeelden/Weinstein-Choice_in_School_Choice:2007} argue that cost is an important barrier to acquiring information. Indirect evidence is provided by \citet{Lovenheim/Walsh-School_Choice_Increase_Information:2018}, who show that an increase in the value of information leads to higher search intensity. \cite{Hastings/Weinstein-Estimating_Impact_on_Choices_Outcomes:2007,Hastings/Weinstein-Information_School_Choice_Achievement_Two_Experiments:2008,Hoxby/Turner_High-achieving_Low-income_Know_about_College:2015,Kessel/Olme_Are_Parents_Uninformed:2017} observe that applicants change their choices when additional information is provided to them, implying that they are not fully informed at the time of the application. A large number of papers report differential information possessed by applicants in different settings or domains \citep{Dur/Hammond/Morrill-Identifying_Harm:2018,Hastings/Neilson/Ramirez/Zimmerman-Uninformed_Choice:2016,Kapor/Neilson/Zimmerman-Heterogeneous_Beliefs_School_Choice:2018,Luflade-Value_of_Information:2018}.
\label{ftnt:Empirical literature}}

It is typically assumed the applicants know their preferences; hence the main question is preference elicitation. However, a truthful uninformed applicant submits a public ranking containing no useful information on her preferences.

Widely advocated mechanisms, such as Deferred Acceptance (DA), elicit preferences truthfully, but if they also discourage learning, their outcomes may be incorrectly assessed. Consider, for example, a school district contemplating a switch from the manipulable Immediate Acceptance (IA) mechanism to DA. To evaluate its effects, the district estimates applicants' preferences by either running a survey, as in \citet{deHaan/Gautier/Oosterbeek/vanderKlaauw-Performance_School_Assignment_Practice:2015} or \citet{Kapor/Neilson/Zimmerman-Heterogeneous_Beliefs_School_Choice:2018}, or using empirical methods that rely on Rank-Ordered Lists (ROLs) submitted to the mechanism, as in \citet{Agarwal/Somaini-Demand_Analysis_Strategic_Reports:2018}, \citet{Calsamglia/Chao/Guell-Structural_Estimation_Boston:2018}, \citet{He-Gaming_Boston_Beijing:2017} or \citet{Hwang_Heterogeneity_Boston_Estimation:2017}. If, following the adoption of DA, applicants start to rely more on public ranking and less on their first-hand experience with schools -- becoming ``uninformed'' in the language of this paper -- they submit more homogeneous ROLs. As a result, increased congestion at more popular schools may hurt applicants for whom these schools are a good fit. The extent of strategizing under IA may also be overstated if ROLs submitted to DA are interpreted as true preferences of fully informed applicants.

I analyze several stylized environments and find that too many applicants are uninformed, under a mild condition, in a strategy-proof random serial dictatorship mechanism. In the mechanism, each applicant is endowed with a ``score'' and assigned to her most preferred ``feasible'' school. A school is feasible if the applicant's score is above a cutoff set by the school.

In all environments I study, I assume that an uninformed applicant does not know applicant-specific school characteristics but knows the public school ranking. A public ranking is a common feature of many environments; an example is US News and World Report rankings of colleges and high schools. The cost of learning a public ranking is negligible; often, it is a single Google search. I assume it is zero. This assumption -- effectively, the common prior -- is crucial for my model, as it leads to homogeneous preferences among uninformed applicants.\footnote{This assumption may not hold in some environments. For example, some families may find academic achievement statistics hard to interpret and base their decisions on other measures such as school proximity \citep{Hastings/Weinstein-Information_School_Choice_Achievement_Two_Experiments:2008}. Uninformed applicants' preferences would then be heterogeneous, and my results would not apply. However, simple school rankings are common. Even for the Charlotte-Mecklenburg School district studied by Hastings and Weinstein, rankings are now compiled in an easy-to-digest format by at least two major ranking websites: Niche and SchoolDigger.}

Any information beyond the public ranking is costly to obtain because it is applicant-specific. I assume that the costs differ among applicants. Such differences may be due to the opportunity cost of time or the kind of information an applicant seeks.

In summary, applicants are ex-ante identical except for their cost of information acquisition. Once they pay the cost, they learn their utilities for all schools. Under these assumptions, both the equilibria and the social optimum are characterized by fractions of informed applicants and can be unambiguously compared. Under a mild condition, the equilibrium fraction of informed applicants is below the social optimum. If equilibrium is not unique, the low-information one is Pareto dominated by the high-information one. 
	
This conclusion is based on (i) a simple model with three schools and symmetric distribution of school quality, presented in the paper, and two extensions, presented in the appendix, (ii) one with a non-symmetric distribution and (iii) one with an arbitrary number of schools but a specific type of the distribution. In all three environments, information is under-acquired under the following condition. Suppose some school $B$ is the highest-ranked school (in the public ranking) that is feasible for applicant $i$. Then the probability that $i$ prefers some other \emph{feasible} school -- such as $C$ -- should be sufficiently low. This condition is imposed independently for each school. It does not change when there are many schools: we are still only interested in the probability that the applicant's \emph{best feasible} school according to the public ranking is not her best feasible school after acquiring information. For example, in the $N$-school environment, the probability that a relative ranking of any given pair of schools differs between individual and public ranking is fixed to be at most $\nicefrac{1}{4}$, but the probability that complete individual and public rankings are identical is close to zero for a large $N$.

The probability that an applicant prefers a feasible school which is not the highest-ranked is ``sufficiently low'' when it is below about $\nicefrac{1}{3}$, depending on the environment. 
We can see that $\nicefrac{1}{3}$ is not very restrictive by continuing with the example above. Note that (a) all uninformed applicants necessarily prefer $B$ to a school ranked lower in the public ranking and (b) if many informed applicants have a low draw for $B$, many others should draw high $B$.\footnote{In all environments, I impose an additional condition on the tails of the distribution, preventing them from being very different, thus avoiding a situation where few very high draws cancel many small low draws for $B$.} In symmetric case, the theoretical maximum is $\nicefrac{1}{2}$, and when it is $\nicefrac{1}{2}$, the public ranking must \emph{never} be ex-post correct. In a model with an arbitrary number of schools and sufficiently similar quotas, the condition trivially holds when other assumptions I impose are satisfied. Thus, we may expect insufficient learning in many realistic environments. Simulations also support that conclusion, identifying only a small number of instances where enough applicants learn.

When applicants learn, they create positive sorting and negative displacement externalities. The sorting externality arises when an applicant, once informed, frees up a seat at a popular school. For example, suppose the public ranking is $ABC$ and consider an applicant, Ann, for whom $A$ is not feasible, but $B$ is. If she learns that her preferences are $ACB$, she frees up a seat at $B$ for another applicant, Bob. The sorting externality Ann creates would arise with few and with many informed applicants. Unlike sorting, displacement only arises when there are many applicants whose preferences differ from $ABC$. To see that, suppose that a high-priority applicant learns that her preferences are $BAC$. She frees up a seat at a popular $A$ and takes a seat at a less-popular $B$. Yet, if Ann takes the seat at $A$, then Bob is displaced from $B$: the externality is positive for Ann but negative for Bob. The negative externality only arises if it is likely that an applicant like Ann takes a seat at $A$, hinting at the condition described in the previous paragraphs.

Further, I study three interventions that may motivate the design of policies that encourage learning. They are simplified versions of existing procedures designed for other aims.
\begin{enumerate}
	\item Applicant's priorities may be released before applicants decide to learn. The policy leads to Pareto improvement but still results in insufficient learning of some groups of applicants.\footnote{The timing of the release of applicants' scores -- either before or after they submit their ROLs -- has been a topic of recent research \citep{Chen/Pereyra-Self-selection_School_Choice:2018,Lien/Zheng/Zhong_Pre-exam_Post_exam_Experiment:2016,Lien/Zheng/Zhong_Pre-exam_Post_exam:2017,Wu/Zhong_Pre-post_Ex-ante_fairness_China:2014}. In most markets, scores are released before applicants submit their ROLs; the authors argue that the delayed release of scores can achieve ex-ante fairness. However, the change in timing also changes incentives to acquire information, and any gains from the late release of scores may be canceled out by applicant's changes in their information acquisition behavior.}
	\item School quotas may be redistributed between informed and uninformed applicants -- in effect, redesigning priorities -- to incentivize learning, leading to a Pareto improvement. A decision to learn must either be directly observed, for example, by collating names of open day visitors\footnote{For example, New York City prioritized applicants attending open days \citep{NYC_school_report}.} or elicited ex-ante.
		
	Affirmative action can be viewed as an instance of this policy, in which prioritization is directed at a specific target group of applicants. The policy can be designed to benefit both target and non-target groups. This is important because affirmative action -- or any priority redistribution -- is often perceived as a zero-sum game in school assignment \citep{Cantillon-Broadening_Approach_to_School_Choice:2017}.
    
	\item A school district can directly subsidize information acquisition, for example, by running information sessions in disadvantaged neighborhoods or translating documents to other languages.\footnote{Similar interventions have been studied in empirical papers cited earlier, e.g., \citet{Hastings/VanWeelden/Weinstein-Choice_in_School_Choice:2007,Hoxby/Turner_High-achieving_Low-income_Know_about_College:2015}} Those services are funded by taxing all applicants. Importantly, I do not assume perfectly transferable utility. I only assume that taxes lead to \emph{some} increase in service/subsidy provided to a specific group of applicants. Then there exists a level of taxation that leads to a Pareto improvement.
\end{enumerate}

These interventions are studied in a simple three-school symmetric environment. My general conclusion on under-investment in learning is also derived for specific environments, and only simulations hint at its wider applicability. A more general model is desirable, but it becomes intractable because the cutoff of a school generally depends on the cutoffs of all the higher-ranked schools (my assumption in the $N$-school model limits that dependence). Even when a closed-form expression can describe an equilibrium, it cannot be analyzed without numerical methods. Thus, simulations allow for greater flexibility in defining the environment without imposing any significant new costs.

Information acquisition in matching markets was first considered by \citet{bade:14}. Bade shows that serial dictatorship provides the best informational incentives among all non-bossy strategy-proof mechanisms. I show that those incentives are still short of socially optimal in most cases and propose policies to improve information acquisition. Two papers also differ in the learning technology they assume. In Bade's setting, the state space is finite, and an applicant learns its partition. I assume that an applicant learns the true state of a continuum state space.
	
Two other, contemporaneous studies differ in both the focus and the setting.
	
\citet{Harless/Manjunath-Learning_Matters:2018} also focus on non-bossy strategy-proof mechanisms. Unlike this paper, they assume that learning is costless; however, applicants can sample only one of multiple uncertain objects. That is, they focus on what to learn rather than whether to learn. Thus, the results of \citet{Harless/Manjunath-Learning_Matters:2018} cannot be directly compared to the results of this paper.
	
\citet{Chen/He-Info_Theory:2018,Chen/He-Info_Experiment:2018} share the motivation but focus on different questions and have some important modeling differences. First, applicants can learn their own and other's preferences in their model; mine does not allow for the latter. Focusing on learning own preferences, all three papers find that IA incentivizes learning better than DA (although my conclusion has caveats). They also compare the welfare of different groups under IA and DA and under different information treatments, which I do not do. In turn, I derive conditions when equilibrium learning is below socially optimal. Technically, I focus on simple environments, either with three schools or with a specific distribution of school valuations. I also have a continuum of applicants. That allows me to make comparisons to the socially optimal level of learning, which their paper lacks. I also assume that uninformed applicants have homogeneous preferences; their theoretical paper mainly focuses on heterogeneous preferences.
	
Interviewing in matching markets \citep{Das/Li-Common_Private_Signals_Matching_Interviewing:2014,Kadam-Interviewing:2015,Lee/Schwarz-Interviewing:2017,Lien_Interviewing:2009} is related to information acquisition, but an assumption that a school must interview an applicant before making an offer creates an incentive that is entirely absent in the information acquisition literature: a school prefers to interview applicants with fewer interviews with other schools, as such applicants would have fewer offers to choose from and are more likely to accept that school's offer. \citet{Drummond/Boutilier_Elicitation_Approximate_Stability:2013,Drummond/Boutilier_Elicitation_Stability:2014} and \citet{Rastegari/Condon/Immorlica/Leyton-Brown-Partial_Information:2013} use similar framework to study an unrelated problem of scheduling costly interviews.
	
A range of papers allow for incomplete information in matching models, but do not allow applicants to search for information \citep{Aziz/Biro/Gaspers/deHaan/Mattei/Rastegari-Matching_Uncertain_Linear:2016,Chade/Lewis/Smith-Portfolios_Matching:2014,Chakraborty/Citanna/Ostrovsky-Matching_Interdependent_Values:2010,Ehlers/Masso-Incomplete_Information_Matching:2015,Liu/Mailath/Postlewaite/Samuelson-Stable_Matching_Incomplete_Information:2014}. Outside of matching theory, a rich literature on information acquisition is surveyed by \citet{Bergemann/Valimaki-Information_Mechanism_Design:2006}. 
	
The rest of the paper is organized as follows. In Section \ref{sec:ex1}, I provide an example to illustrate the positive and negative externalities of learning. I describe the model in Section \ref{sec:setup}, find equilibria in Section \ref{sec:DAA}, and obtain a condition for under-acquisition of information in Section \ref{sec:SO}. Example in Section \ref{sec:exmpls} demonstrates that this condition is tight. Section \ref{sec:policy} describes Pareto-improving interventions. In Section \ref{sec:IA}, I compare learning incentives to the Immediate Acceptance mechanism. In Section \ref{sec:Simulations}, I provide simulations for a richer model and discuss extensions in Section \ref{sec:Discussion}. I conclude in Section \ref{sec:conclusion}.
	
\section{An example}\label{sec:ex1}

Example \ref{ex1} illustrates externalities resulting from one applicant's decision to acquire information. When only a few applicants are informed, the externality is necessarily positive. As the applicant deciding to learn does not take the positive externality into account, she may opt against learning even when it is socially optimal. When nearly everyone is informed, the effect is more nuanced: the uninformed are still positively affected by learning, but the informed may not be.

\begin{example}\label{ex1}
	Consider a problem in which five applicants, $\mathbf{1},$ $\mathbf{2},$ $\mathbf{3},\mathbf{4}$ and $\mathbf{5}$ are allocated to three schools, $A,$ $B$ and $C$. Schools $A$ and $B$ have one seat each, and school $C$ has three seats. Each applicant finds every school acceptable.
	
	For each applicant $i \in \{\mathbf{1},\mathbf{2},\mathbf{3},\mathbf{4},\mathbf{5}\}$, $i$'s utilities are: $u_i (A) = 1,$ $u_i (C) = 0$ and $u_i(B) = 1.5$ or $u_i(B) = -1$ with probability $\nicefrac{1}{2}$ each.
	
	The applicants are allocated using Random Serial Dictatorship (RSD): they are randomly assigned priorities, 1, 2, 3, 4, and 5, then each selects the most desirable school that is still available, in the order of their priority. Since RSD is a strategy-proof mechanism, applicants submit their preferences truthfully. An uninformed applicant's ROL is $ABC$, while an informed applicant's ROL could be either $BAC$ or $ACB$ with equal probability.

	If no applicant is informed, then their probability of being assigned to a given school is identical and shown in Table \ref{Example1Table0}.

  \begin{table}[h]
    \centering
    \begin{tabular}{|r|c|c|c|c|}
			\cline{3-5}
			\multicolumn{2}{c|}{}&\multicolumn{3}{|c|}{Schools}\\
      \cline{3-5}
      \multicolumn{2}{c|}{} & $A$   & $B$   & $C$\\
      \hline
      \multicolumn{2}{|c|}{Applicants $\mathbf{1},\mathbf{2},\mathbf{3},\mathbf{4}, \mathbf{5}$}   & $\nicefrac{1}{5}$   & $\nicefrac{1}{5}$   & $\nicefrac{3}{5}$ \\
      \hline
    \end{tabular}
    \caption{Assignment probabilities when $R_{\mathbf{1},\mathbf{2},\mathbf{3},\mathbf{4},\mathbf{5}}=ABC$.}
    \label{Example1Table0}
  \end{table}
	Suppose now that $\mathbf{1}$ learns that $B$ is her top choice and submits $BAC$. All other applicants submit $ABC$. The probabilities of assignments change to those given in Table \ref{Example1Table2}.
  \begin{table}[h]
    \centering
    \begin{tabular}{|rc|c|c|c|}
			\cline{3-5}
			\multicolumn{2}{c|}{}&\multicolumn{3}{|c|}{Schools}\\
      \cline{3-5}
      \multicolumn{2}{c|}{} & $A$   & $B$   & $C$\\
      \hline
			\multirow{2}{*}{Applicants}
      &$\mathbf{1}$       			& 0     & $\nicefrac{2}{5}$   & $\nicefrac{3}{5}$ \\
      &$\mathbf{2}, \mathbf{3}, \mathbf{4}, \mathbf{5}$	& $\nicefrac{1}{4}$   & $\nicefrac{3}{20}$   & $\nicefrac{3}{5}$ \\
      \hline
    \end{tabular}
    \caption{Assignment probabilities when $R_\mathbf{1}=BAC$ and $R_{\mathbf{2},\mathbf{3},\mathbf{4}, \mathbf{5}}=ABC$.}
    \label{Example1Table2}
  \end{table}
	Note that $\mathbf{1}$ would be assigned to school $B$ if her lottery priority is either 1 or 2 and would be assigned to $C$ if her priority is 3, 4 or 5. All other applicants do not face any competition for $A$ from applicant $\mathbf{1}$; their probability of an assignment to $A$ increases to $\nicefrac{1}{4}$. Indeed, $i \in \{\mathbf{2},\mathbf{3},\mathbf{4},\mathbf{5}\}$ is allocated to $A$ if his priority is 1 (probability is $\nicefrac{1}{5}$), or if his priority is 2 and $\mathbf{1}$'s priority is 1 (probability is $\nicefrac{1}{20}$). All other probabilities can be similarly calculated.

	Note that $i \in \{\mathbf{2},\mathbf{3},\mathbf{4}, \mathbf{5}\}$ are better off after $\mathbf{1}$ learns: although the total probability of being allocated to either $A$ or $B$ remains the same, the probability of being allocated to the more preferred school $A$ increases. Thus, the decision of $\mathbf{1}$ to learn creates a positive externality on other applicants. I call this positive externality \emph{sorting} because it leads to a more efficient sorting of applicants to schools, thus increasing their expected utility.

	Suppose next that $\mathbf{1}, \mathbf{2}$ and $\mathbf{3}$ are informed; $\mathbf{1}$ learns that $B$ is preferred to $C$ and submits $R_\mathbf{1}=BAC$ and both $\mathbf{2}$ and $\mathbf{3}$ learn that $C$ is preferred to $B$ and submit $R_{\mathbf{2},\mathbf{3}} = ACB$. $\mathbf{4}$ and $\mathbf{5}$ remain uninformed. Then the assignment probabilities are given in Table \ref{Example1Table3}. Note that all applicants are better off compared to Table \ref{Example1Table2}.
  \begin{table}[h]
    \centering
    \begin{tabular}{|rc|c|c|c|}
			\cline{3-5}
			\multicolumn{2}{c|}{}&\multicolumn{3}{|c|}{Schools}\\
      \cline{3-5}
      \multicolumn{2}{c|}{} & $A$   & $B$   & $C$\\
      \hline
			\multirow{3}{*}{Applicants}
       &$\mathbf{1}$       				& 0     & $\nicefrac{16}{30}$   & $\nicefrac{14}{30}$ \\
       &$\mathbf{2}, \mathbf{3}$  & $\nicefrac{1}{4}$   & 0     & $\nicefrac{3}{4}$ \\
       &$\mathbf{4}, \mathbf{5}$    & $\nicefrac{1}{4}$   & $\nicefrac{7}{30}$   & $\nicefrac{31}{60}$ \\
      \hline
    \end{tabular}
    \caption{Assignment probabilities when $R_\mathbf{1}=BAC$, $R_{\mathbf{2},\mathbf{3}} = ACB$ and $R_{\mathbf{4}, \mathbf{5}}=ABC$.}
    \label{Example1Table3}
  \end{table}

    Finally, suppose that $\mathbf{4}$ learns that $B$ is better than $A$ and submits $R_\mathbf{4} = BAC$. Then the probabilities become as in Table \ref{Example1Table4}. Note that applicant $\mathbf{1}$ is worse off when $\mathbf{4}$ acquires information; the decision of $\mathbf{4}$ to learn creates a negative \emph{displacement} externality. This might be surprising because $\mathbf{4}$ ``exchanges'' a valuable seat at $A$ for a seat at $B$; why does it make $\mathbf{1}$ worse off? If it were true that the seat is simply exchanged (e.g. with applicant $\mathbf{5}$), then there would be no displacement externality. However, a seat at $A$ may be taken by $\mathbf{2}$ or $\mathbf{3}$, who, before $\mathbf{4}$ learns, are assigned to $C$ because their priority is too low for a seat at $A$. 
	
  \begin{table}[h]
    \centering
    \begin{tabular}{|rc|c|c|c|}
			\cline{3-5}
			\multicolumn{2}{c|}{}&\multicolumn{3}{|c|}{Schools}\\
      \cline{3-5}
      \multicolumn{2}{c|}{} & $A$   & $B$   & $C$\\
      \hline
			\multirow{2}{*}{Applicants}
      &$\mathbf{1}, \mathbf{4}$ & $\nicefrac{1}{20}$  & $\nicefrac{13}{30}$   & $\nicefrac{31}{60}$ \\
      &$\mathbf{2}, \mathbf{3}$ & $\nicefrac{3}{10}$  & 0     & $\nicefrac{7}{10}$ \\
      &$\mathbf{5}$ & $\nicefrac{3}{10}$  & $\nicefrac{2}{15}$     & $\nicefrac{17}{30}$ \\
      \hline
    \end{tabular}
    \caption{Assignment probabilities when $R_{\mathbf{1},\mathbf{4}}=BAC$, $R_{\mathbf{2},\mathbf{3}}=ACB$ and $R_\mathbf{5}=ABC$.}
    \label{Example1Table4}
  \end{table}

	It is intuitive, and I will verify it formally below, that uninformed applicants are always better off when other applicants acquire information. The displacement externality is negative and may outweigh sorting only for informed applicants and only when many applicants are informed.

\end{example}

\section{Setup}\label{sec:setup} 

The paper follows the continuum matching framework introduced by \cite{azevedo/leshno:16}, who show that assignments in large finite economies converge to the assignment in the continuum economy. I present a model with three schools and symmetric distribution in the main body of the paper. The main result that establishes a condition for under-acquisition of information is shown for two extensions, with an asymmetric distribution (Appendix \ref{sec:appendix2}) and with an arbitrary number of schools but specific assumption on the distribution (Appendix \ref{subsec:appendix3}).

A mass 1 of applicants is allocated to three schools, $A$, $B$, and $C$. The quotas in schools $A$ and $B$ are identical and equal to $q \leq \nicefrac{1}{3}$. The quota at school $C$ is 1. That is, it can accept all applicants not assigned to schools $A$ and $B$ and can be thought of as an outside option.

The utilities of applicants are\footnote{
This specification fixes the order of $A$ and $C$. If I were to allow the reversal, it would increase the positive externality from learning (because of the outflow of applicants from $A$ and $B$) without introducing any new negative externalities; thus, learning would be more socially desirable.
}
\begin{gather*}
	u_i(A) = 1;  \qquad u_i(B) = \frac{1}{2} + \epsilon_i; \qquad u_i(C) = 0,
\end{gather*}
where $\epsilon_i$ is drawn from distribution $\mathcal{F}(x)$ with zero mean, symmetric around zero and with no atoms at $-\nicefrac{1}{2}$ and $\nicefrac{1}{2}$, implying that an applicant is indifferent between two schools with probability zero. The distribution is identical for all the applicants. Applicant $i$ does not know the realization of $\epsilon_i$ but can learn it at cost $c(i)$. Applicants are identical ex-ante except for this cost. I assume that $c(i)$ is strictly increasing. Note that assuming an increasing $c(i)$ is without loss of generality. Assuming that it is \emph{strictly} increasing simplifies the argument because the inverse of $c(i)$ exists.

Denote the probability that $\epsilon_i > 1/2$ or $\epsilon_i < -1/2$ -- that is, the probability that the ordering of $A$ and $B$ or of $B$ and $C$ changes -- by
\begin{gather}
	p = \mathcal{F}\left(-\frac{1}{2}\right) = 1 - \mathcal{F}\left(\frac{1}{2}\right),\label{def:Phi}
\end{gather}
Suppose that an applicant learns $\epsilon_i$, her ordinal ranking changes, and she is assigned to the school she prefers instead of the school which is better in expectation. Denote the gain from this assignment by
\begin{gather}
	\mathcal{G} = E\left[\epsilon_i-\frac{1}{2}\big|\epsilon_i > \frac{1}{2}\right] = -E\left[\epsilon_i+\frac{1}{2}\big|\epsilon_i<-\frac{1}{2}\right]
	\label{def:I}
\end{gather}

The timing is as follows. First, applicants learn their disutility from acquiring information, $c(i)$, which can be assumed to be drawn from an arbitrary atomless continuous distribution. After observing the cost, some applicants choose to learn $\epsilon_i$. Then all applicants submit their ROLs to the centralized clearinghouse without observing anything about other applicants\footnote{The assumption that applicants do not observe the decisions of the others simplifies the notation but can be easily dispensed with. Indeed, applicants base their decisions on aggregates, which are known in equilibrium and are not affected by individual deviations.} and are assigned to schools. That is, applicant's strategy is $s_i=(e_i,\hat{R}_i(\epsilon_i))$, where $e_i \in \{0,1\}$ denotes the learning decision and $\hat{R}_i$, applicant's ROL, is a permutation of $A, B$ and $C$, which may depend on the realization of $\epsilon_i$.

An applicant's priority is determined by a (uniformly distributed) random number $r_i \in [0,1]$ so that if $r_i < r_j$, then applicant $i$ has higher priority than $j$ at every school.\footnote{The tie-breaking rule that gives applicants the same priorities at all schools is called ``single tie-breaking.'' It has efficiency benefits over other rules \citep{pathak/sethuraman:09}.} In the main part, I assume that applicants do not know their priorities when they decide to acquire information; other cases are handled in Sections \ref{sc:revealing} and \ref{sec:priorities}.

Since applicant $i$'s random priority $r_i$ is the same at all schools, DA is equivalent to RSD.

\section{Equilibria in Random Serial Dictatorship}\label{sec:DAA}

In this section, I find equilibria in the game where applicants decide whether to acquire information and are assigned using RSD. Each equilibrium is characterized by the fraction of applicants who learn, $\gamma^{Eq}$: applicants with cost $c(i) \leq c(\gamma^{Eq})$ learn, and the rest of the applicants do not.

I first establish that it is sufficient to focus on applicants' truthful reporting. Under truthful reporting and for a given share of informed applicants $\gamma$, equating supply of and demand for seats allow me to find cutoffs -- the minimum score an applicant must have to be accepted to a school -- of schools $A$ and $B$, denoted $r(A;\gamma)$ and $r(B;\gamma)$. The assumptions on the environment guarantee that $A$ is more selective than $B$, $r(A;\gamma) < r(B;\gamma)$,\footnote{Note that lower score $r_i$ is better; this definition of $r_i$ will later allow me to interpret $r(A;\gamma)$ and $r(B;\gamma)$ as the probability of being assigned to $A$ and $B$, respectively.} for any level of information acquisition $\gamma \in [0,1]$. Thus $r(A;\gamma)$ does not depend on $r(B;\gamma)$. Taking cutoffs as given, I calculate the expected utilities of informed and uninformed applicants. The difference between the two -- the gain from learning -- is equal to the cost of a marginal applicant in the equilibrium in which the fraction of informed applicants is $\gamma^{Eq}$.

The supply and demand equations can be written as follows:
\begin{align}
	q &= r(A;\gamma) \left((1-\gamma) + \gamma(1-p)\right), 
	\label{eq:general-rA}
	\\
 	q &= r(B;\gamma)\gamma p + (r(B;\gamma)-r(A;\gamma))\big((1-\gamma)+\gamma(1 - 2 p)\big).
	\label{eq:r_B_general_v}
\end{align}
The expression in brackets on the RHS of equation (\ref{eq:general-rA}) is the mass of applicants who rank $A$ as their top choice. The first term on the RHS of equation (\ref{eq:r_B_general_v}) captures applicants who rank $B$ first and the second term captures applicants rejected from $A$ whose second choice is $B$.

Equation (\ref{eq:r_B_general_v}) can be re-written, using (\ref{eq:general-rA}), as
\begin{gather}\label{eq:intuition1}
	2 q = r(B;\gamma) (1 - \gamma p) + r(A;\gamma) \gamma p.
\end{gather}
The effect of $\gamma$ on $r(A;\gamma)$ is unambiguous: when there are more informed applicants, there are more applicants who prefer $B$ to $A$ and the competition for seats at $A$ is lower. The RHS of (\ref{eq:intuition1}) shows two effects of $\gamma$ on $r(B;\gamma)$. The first term captures the effect identical to the effect of $\gamma$ on $r(A;\gamma)$: some applicants leave $B$ for $C$. The second term captures an increased demand for $B$ from applicants for whom $A$ is not their first choice. The interplay between these two terms will help me define sorting and displacement externalities. 

The supply and demand equations can be solved to obtain cutoffs
\begin{align}
	r(A;\gamma) &= q\frac{1}{1 - \gamma p}, \label{rA}\\
	r(B;\gamma)	&= q\frac{2 - 3\gamma p}{(1 - \gamma p)^2}.\label{rB}
\end{align}
Note that $\frac{\partial}{\partial \gamma}r(A;\gamma) > 0$ for any $\gamma \in [0,1]$, but $\frac{\partial}{\partial \gamma}r(B;\gamma) > 0$ only when $\gamma p < \nicefrac{1}{3}$.

Ex-ante (before learning $\epsilon_i$) expected utilities of uninformed and informed applicants are
\begin{align}
	U&(0,ABC|r(A;\gamma),r(B;\gamma)) = r(A;\gamma) \times 1 + \big(r(B;\gamma) - r(A;\gamma)\big) \times \frac{1}{2}\label{eq:U(0|gamma)}\\
	U&(1,R(\cdot)|r(A;\gamma),r(B;\gamma)) = 
	\begin{aligned}[t]
	&U(0,ABC|r(A;\gamma),r(B;\gamma))\\
	&+p\ r(A;\gamma)\mathcal{G} + p \ \big(r(B;\gamma) - r(A;\gamma)\big)\mathcal{G},\label{eq:U(1|gamma)}
	\end{aligned}
\end{align}
where the second term on the RHS of (\ref{eq:U(1|gamma)}) is the expected gain when $r_i \leq r(A;\gamma)$ and the third is the expected gain when $r(A;\gamma) < r_i \leq r(B;\gamma)$. Note that (\ref{eq:U(1|gamma)}) does not account for the cost of information acquisition. Expression (\ref{eq:U(1|gamma)}) can be re-written, using (\ref{eq:intuition1}) as 
\begin{align}
    U(1,R(\cdot)&|r(A;\gamma),r(B;\gamma)) = r(A;\gamma)/2 + r(B;\gamma) (1/2 + p \mathcal{G})\label{eq:U(1|gamma)-ver0}
    \\
    &= \underbrace{\frac{q}{1 - \gamma p}\frac{1}{2} + \frac{2q}{1 - \gamma p}(1 + 2 p \mathcal{G})}_{\text{sorting}} - \underbrace{\frac{\gamma p}{1 - \gamma p}\frac{q}{1 - \gamma p} (1 + 2 p \mathcal{G})}_{\text{displacement}}.\label{eq:U(1|gamma)-ver}
\end{align}
When $\gamma$ changes, the changes in the first two terms define positive sorting and the change in the last term defines negative displacement externalities, as experienced by an informed applicant. These definitions extend to environments with three schools and a general distribution $\mathcal{F}(x)$ (equations \ref{eq:sorting-general} and \ref{eq:displacement-general} in Appendix~\ref{sec:appendix2}) and with $N$-schools and the distribution that restricts the outflow of applicants to the next-best school only (equations~\ref{eq:Nschool-r(s_k)} and \ref{eq:multi-util-diff-term} in Appendix~\ref{subsec:appendix3}). The externalities on uninformed applicants can be defined analogously, as changes in two terms in the expression $U(0,\hat R|r(A;\gamma),r(B;\gamma)) =
\frac{q}{2}\left(\frac{3}{1-\gamma p} - \frac{\gamma p}{(1-\gamma p)^2}\right)$. I focus on externalities on informed because, for uninformed, sorting always dominates displacement in all environments I study.

The externalities can be illustrated as follows. If $A$ is feasible for an applicant, she would be assigned to $A$ if she is uninformed, but to $B$ if she learns that her preferences are $BAC$. Hence, when she learns, $A$ becomes less selective; $r(A;\gamma)$ increases. Similarly, when an applicant with preferences $ACB$ learns, she gives up a seat at $B$ if $A$ is not feasible for her but $B$ is. These two observations mean that the first two terms in (\ref{eq:U(1|gamma)-ver}) increase with $\gamma$. This is a positive sorting externality. However, an increase in $r(A;\gamma)$ also leads to a negative displacement externality. To see this effect, consider again an applicant with preferences $ACB$ for whom $A$ is not feasible. She is assigned to $C$. As other applicants learn and $r(A;\gamma)$ increases, $A$ may become feasible for her. She is now assigned to $A$. If all other applicants were to keep their assignments, the total number of applicants at schools $A$ and $B$ would exceed their total capacity. Hence, the cutoff $r(B;\gamma)$ must decrease; this is captured by the second term in (\ref{eq:intuition1}). The corresponding decrease in the utility of informed applicants is captured by the third term in (\ref{eq:U(1|gamma)-ver}). Yet, if applicants with preferences $ACB$ are rare, the probability that a vacated seat is occupied by them is low and soritng dominates displacement.

An environment with asymmetric $\mathcal{F}(x)$ illustrates that the displacement effect may be absent. When there are no applicants with preferences $ACB$ ($p_C = 0$), there are no applicants who move from $C$ to $A$ after a small increase in $r(A;\gamma)$. The preceding argument then implies that there should be no displacement externality and, indeed, the term~(\ref{eq:displacement-general}), corresponding to the displacement externality, is zero.  

In the environment with $N$ schools, an increase in one school's cutoff similarly decreases the cutoff of the school ranked directly below it (equation~\ref{eq:Nschool-r(s_k)}). Unlike the three-school environment, there are additional ripple effects on the cutoffs of all lower-ranked schools. It is possible to account for them because the effects diminish and do not accumulate. For example, when schools have similar quotas, an increase in the cutoff of the top-ranked school leads to a smaller decrease in the cutoff of the second-ranked school, which, in turn, leads to an even smaller \emph{increase} in the cutoff of the third-ranked school, and so on.

With no restrictions on the distribution, $\mathcal{G}$ can be arbitrarily large and whether sorting or displacement externality dominates would primarily depend on the second and third terms of (\ref{eq:U(1|gamma)-ver}), which are related to the change in $r(B;\gamma)$. When $\gamma p > \nicefrac{1}{3}$, $r(B;\gamma)$ decreases in $\gamma$ and a displacement externality may dominate sorting. For that reason, the condition $\gamma p \leq \nicefrac{1}{3}$ features in a number of theorems below. To evaluate the condition $\gamma p \leq \nicefrac{1}{3}$, note that even when the rankings $ABC, BAC$, and $ACB$ are equally likely, $p = \nicefrac{1}{3}$ and the condition is satisfied for any $\gamma$.

The expected gain from learning is defined as the difference between $U(1,R(\cdot)|r(A;\gamma),r(B;\gamma))$ and $U(0,ABC|r(A;\gamma),r(B;\gamma))$ and can be expressed as follows:
\begin{align}
	\Delta U(\gamma) = p \ r(B;\gamma) \mathcal{G}.\label{eq:diffutil}
\end{align}

To establish the existence of equilibrium, consider a function $i^*$, defined as follows:
\begin{align}
i^*(x) = 
	\begin{cases}
		0, &\text{if } x < c(0)\\
		c^{-1}(x), &\text{if } x \in [c(0),c(1)],\\
		1, &\text{if } x > c(1).
	\end{cases}
\end{align}
This function is well-defined because $c(i)$ is strictly increasing, and it is continuous because $c(i)$ is continuous. For any $i^* \in (0,1)$, the function identifies an individual whose benefit of acquiring information is equal to the cost $c(i^*)$.

\begin{definition}
	Let $\Gamma$ be a collection of $\gamma \in [0,1]$ such that:
	\[
		\gamma = i^*\left(\Delta U(\gamma)\right)
	\]
\end{definition}

\begin{theorem}\label{th:Gamma_nonempty}
	The set $\Gamma$ is non-empty.
\end{theorem}

\begin{theorem}\label{th:eqSD}
	For each $\gamma^{Eq} \in \Gamma$, a strategy profile $\{(\hat e_i, \hat R_i)\}_{i \in N}$ defined as:
			\begin{equation}\label{th_eq:RD_e_i}
					\hat e_i=
					\begin{cases}
						0 & \text{if } i > \gamma^{Eq}\\
						1 & \mbox{if } i \leq \gamma^{Eq},
					\end{cases}
			\end{equation}
			and 
			\begin{equation}\label{th_eq:RD_R_i}
					\hat R_i =
					\begin{cases}
						BAC & \mbox{if } \hat e_i = 1 \text{ and } u_i(B) \geq 1\\
						ACB & \mbox{if } \hat e_i = 1 \text{ and } u_i(B) \leq 0\\
						ABC & \mbox{otherwise,}
					\end{cases}
			\end{equation}
	is a Nash equilibrium. 
	
	Furthermore, if a strategy profile $\{(e_i', R_i')\}_{i \in N}$ is a Nash equilibrium, then (i) the fraction of informed applicants in this equilibrium, $\gamma'$, is such that $\gamma' \in \Gamma$ and (ii) for all applicants $i \neq \gamma'$, the assignment is the same as under $\{(\hat e_i, \hat R_i)\}_{i \in N}$, where $\hat e_i$ is defined by (\ref{th_eq:RD_e_i}) for $\gamma^{Eq} = \gamma'$ and $\hat R_i$ is defined by (\ref{th_eq:RD_R_i}).
\end{theorem}

Theorem~\ref{th:eqSD} says that (\ref{th_eq:RD_e_i})--(\ref{th_eq:RD_R_i}) describe essentially all equilibria: all applicants except possibly $i = \gamma'$ make the same information decisions as one of the equilibria described by (\ref{th_eq:RD_e_i})--(\ref{th_eq:RD_R_i}) and submit ROLs that lead to the same assignment as the truthful ROLs. In certain cases, equilibria described in Theorem \ref{th:eqSD} can be Pareto ranked.

\begin{theorem}\label{th:Pareto_ranking}
    Suppose that $\Gamma$ has at least two elements, $\gamma_L$ and $\gamma_H$, such that $\gamma_L < \gamma_H$. If $\gamma_H p \leq \nicefrac{1}{3}$, then the equilibrium associated with $\gamma_H$ Pareto dominates the equilibrium associated with $\gamma_L$.
\end{theorem}

\section{Socially optimal information acquisition under RSD}\label{sec:SO}

I assume that the mechanism -- RSD -- is fixed. The school district directs some applicants to acquire information to maximize total welfare; all applicants willingly submit truthful ROLs. Solving this problem, I show that there is an under-acquisition of information in equilibrium.

	Since $c(i)$ is increasing but applicants are otherwise ex-ante identical, the district's problem can be expressed as selecting $\gamma^{SO} \in [0,1]$, so that applicants $i \leq \gamma^{SO}$ learn and $i > \gamma^{SO}$ do not.\footnote{A more general formulation, where the district selects a function $e: [0,1] \rightarrow \{0,1\}$ so that applicant $i$ acquires information if $e(i) = 1$ and does not if $e(i) = 0$, can straightforwardly be reduced to the case of selecting a single value $\gamma \in [0,1]$.} The value $\gamma^{SO}$ maximizes
\begin{equation}
	SW(\gamma) = (1-\gamma)U(0,ABC|\gamma) + \gamma U(1,\hat{R}|\gamma) - \int_0^{\gamma} c(i)di \label{eq:soc_welf_int_3},
\end{equation}
	where the dependence of expected utility on $r(A;\gamma)$ and $r(B;\gamma)$ is expressed as dependence on $\gamma$. The expression (\ref{eq:soc_welf_int_3}) can be interpreted as the ex-ante -- before $i$ knows her cost realization $c(i)$ -- expected utility of applicant $i$. Indeed, with probability $(1-\gamma)$, $i > \gamma$, so $i$ is uninformed and obtains $U(0,ABC|\gamma)$. With probability $\gamma$, $i \leq \gamma$, so $i$ is informed and obtains $U(1,\hat R|\gamma)$ paying, in expectation, the cost $\int_0^\gamma c(i) di$. Thus, $\gamma^{SO}$ maximizes ex-ante expected utility of applicant $i$.
Differentiating with respect to $\gamma$, I obtain
\begin{align}
	\frac{\partial}{\partial \gamma} SW(\gamma) 
	= \big[\Delta U(\gamma) - c(\gamma)\big] + \gamma \frac{\partial}{\partial \gamma} U(1,R(\cdot)|\gamma) + (1-\gamma) \frac{\partial}{\partial \gamma}U(0,ABC|\gamma)\label{eq:dSW}
\end{align}
Expression (\ref{eq:dSW}) leads to the following theorem:
\begin{theorem}\label{th:SO-more-than-DA}
    For any $\gamma^{Eq} \in \Gamma$, if $\gamma^{Eq} p \leq \nicefrac{1}{3}$ and $\gamma^{Eq} < 1$, then $\gamma^{Eq} < \gamma^{SO}$.
\end{theorem}
Note that at $\gamma^{Eq} \in \Gamma$, $\Delta U(\gamma^{Eq}) - c(\gamma^{Eq})$ is equal to zero. As noted earlier, sorting (weakly) dominates displacement for informed applicants when $\gamma p \leq \nicefrac{1}{3}$ and for uninformed applicants for any $\gamma \in [0,1]$. Thus, $\gamma p \leq \nicefrac{1}{3}$ guarantees that there is under-acquisition of information. On the other hand, absent restrictions on $\mathcal{F}(x)$, a displacement externality for informed applicants may dominate sorting when $\gamma p > \nicefrac{1}{3}$, as the next example shows.

\section{Example: tightness of $p \leq \nicefrac{1}{3}$ condition}\label{sec:exmpls}

In the example below I show that if the condition $p \leq \nicefrac{1}{3}$ is violated, then there exists a distribution $\mathcal{F}(x)$ and a cost function $c(i)$ such that $\gamma^{Eq} > \gamma^{SO}$. In other words, if there are no additional restrictions on $\mathcal{F}(x)$ and $c(i)$, then the condition in Theorem \ref{th:SO-more-than-DA} cannot be improved.

I argued earlier that $\mathcal{G}$ can be made sufficiently large so that only the changes in $r(B;\gamma)$ determine whether the information is under- or over-acquired. In the example, I construct a distribution which puts the probability just over $\nicefrac{1}{3}$ on a sufficiently high realization of $\epsilon_i$. I then select a cost function so that $i = 1$ is just indifferent to learning and thus $\gamma^{Eq} = 1$.

\begin{example}\label{Ex:RD_SW2}

Fix an arbitrarily small $\phi > 0$. Let $p = (1+\phi)/3$. Note that $r(A;\gamma)$ and $r(B;\gamma)$ depends on $p$ but not any other detail of $\mathcal{F}(x)$. Thus, once we fixed $p$, the following quantity is well-defined:
\begin{align}
	\bar{x} = \frac{3}{2(1+\phi)}\frac{\frac{\partial}{\partial \gamma}\left(r(A;1)+r(B;1)\right)}{\frac{\partial}{\partial \gamma}r(B;1)} < 0,
\end{align}
where the last inequality holds for any $\phi > 0$. This value decreases with $\phi$; for example, for $\phi=0.1$, $\bar{x} \approx -7$ and for $\phi=0.01$, $\bar{x} \approx - 97$. Pick $x_0 < \bar{x}$ and define the distribution as follows:
	\begin{equation*}
		\mathcal{F}(x) = 
		\begin{cases}
			0 & \text{ for } x < x_0;\\
			(1+\phi)/3 & \text{ for } x \in [x_0,- 1/2];\\
			1/2 + x \left(1 - 2(1+\phi)/3\right) & \text{ for } x \in [-1/2,1/2];\\
			1 - (1+\phi)/3 & \text{ for } x \in [1/2,- x_0];\\
			1 & \text{ for } x > - x_0.			
		\end{cases}
	\end{equation*}
For this distribution, $p \mathcal{G} = |x_0|(1+\phi)/3$. Letting
\[
	c(i) = \frac{1 - \phi^2}{2 - \phi^2} p \mathcal{G} \times i
\]
we guarantee $\gamma^{Eq} = 1$.

At $\gamma = 1$, equation (\ref{eq:dSW}) can be written as 
	\begin{align}
		\frac{\partial}{\partial \gamma} SW(1) =  p \mathcal{G} \frac{\partial}{\partial \gamma}r(B;1) + \frac{1}{2}\frac{\partial}{\partial \gamma}\big(r(A;1)+r(B;1)\big) < 0,\label{ex:SW_example}
	\end{align}
	where the last inequality follows from the choice of $x_0$. Thus, $1 = \gamma^{Eq} > \gamma^{SO}$.
\end{example}

\section{Interventions leading to Pareto improvements}\label{sec:policy}

    In this section, I introduce three interventions that improve incentives to learn and lead to Pareto improvement. Those interventions resemble realistic policies and may serve as a motivation for exploring the effect of these policies on learning. Throughout the section, the cost is implicitly chosen so that $\gamma^{Eq} < 1$.

    \subsection{Revealing lottery priorities}\label{sc:revealing}

	I consider the same model as in Section \ref{sec:DAA}, with the exception that applicants know their priorities, $r_i$, before they learn; thus, they base their learning decisions on both the priority and the cost.\footnote{The policies of full information revelation and non-revelation can be seen as two extremes of a more general policy, where the designer reveals priorities only to some applicants. However, in this environment, it is always (weakly) optimal to reveal priority information to all applicants.} I continue to assume that the lottery that determines applicant's priority is not correlated with her learning cost. I show that the intervention leads to a Pareto improvement (Theorem~\ref{th:Pareto_improvement_revelation}), but the under-acquisition of information persists (Theorem~\ref{th:revealing_ranks_underacquisition}).
			
    \begin{theorem}\label{th:Pareto_improvement_revelation}
        The utilities of applicants weakly increase when priorities are revealed before information acquisition. There is a positive mass of applicants for whom the increase is strict.
    \end{theorem}
    
    Revealing priorities removes one element of uncertainty about the gain from learning. Learning by a low-priority applicant, for whom the only feasible school in equilibrium is $C$, is wasteful. Once they know their priorities, they do not learn. Applicants who are not low priority know with certainty that, for one realization of $\epsilon_i$, they can improve their assignment if they learn; effectively, $p=1$ in equation (\ref{eq:diffutil}) for them. One possible complication is that, as applicants change their learning decisions, cutoffs also change; for example, in some environments, the total mass of informed applicants may decrease after priorities are revealed. Despite this, the cutoffs unambiguously increase because any possible decline in the total mass of informed applicants comes from low-priority applicants, whose learning does not affect cutoffs. The increase in $r(B;\gamma)$ further increases the gain from learning ($\mathcal{G}$) for medium- and high-priority applicants, whose equilibrium feasible schools are $B,C$ and $A,B,C$, respectively.
    
    Despite the Pareto improvement, there are still groups of applicants who under-acquire information. I assume that, since applicants' priorities are known, the social planner can set different levels of information acquisition for three different groups (low-, medium- and high-priority applicants). As in the main model, the decision to learn by medium- and high-priority applicants lead to a positive sorting externality. For medium-priority applicants, there is no countervailing displacement externality, as $A$ is not feasible for them: their assignment changes only when they learn $ACB$ and reduce the demand for school $B$. There is no change, and no externality, when they learn $BAC$. Hence medium-priority applicants unconditionally under-acquire information. High-priority applicants may also under-acquire information under certain conditions, for reasons essentially identical to those in the environment with unknown priorities.
    
	\begin{theorem}\label{th:revealing_ranks_underacquisition}
		When priorities are revealed before information acquisition, there exist two thresholds on applicant scores $0 < r_{\pmb a} < r_{\pmb b} < 1$ such that
		\begin{enumerate}
			\item In the group of applicants with $r_i < r_{\pmb a}$, too few are informed if the gain $\mathcal{G}$ is not much higher than the cost $c(i)$ (unless all applicants in that group are informed) and too many otherwise (unless no applicants in the group are informed);
			\item In the group of applicants with $r_{\pmb a} < r_i \leq r_{\pmb b}$, too few are informed (unless all applicants in that group are informed);
			\item Applicants with priorities $r_i > r_{\pmb b}$ are uninformed and it is socially optimal.
		\end{enumerate}
	\end{theorem}
	The statement that the gain $\mathcal{G}$ is not much higher than the cost $c(i)$ is made precise in Appendix \ref{sec:appendix2}, but both normal and uniform distributions satisfy that condition as long as $p \leq \nicefrac{1}{3}$.

\subsection{Redesigning priorities}
\label{sec:redesign_priority}

In this section, I assume that the district can observe the information decision and can treat informed and uninformed groups differently. The district would allocate relatively more seats in $B$ and relatively fewer seats in $A$ to the informed than to the uninformed group. Alternatively, this policy can be thought of as redesigning groups' priorities for $A$ and $B$. This intervention is Pareto improving because informed applicants value seats at $B$ more than uninformed applicants: depending on the information obtained, an informed applicant may gain by choosing $B$ instead of $A$ or $C$ instead of $B$. Uninformed applicants may be compensated by the loss of seats in $B$ by seats in $A$.

The district may be able to observe the information decision directly if it is the primary provider of information (e.g., by collating the names of open day attendees). It may elicit an ex-ante intention to learn in other ways. It is ex-ante incentive-compatible for applicants to report their intention due to the very design of the scheme: for example, if an informed applicant pretends to be uninformed, she has a lower chance to be allocated a seat at $B$, which, in turn, lowers her chance to act upon her information; hence the value of information is lower. It is essential, however, that an applicant reports her intention before learning: an applicant who learns that $B$ is worse than $C$ would be better off if grouped with uninformed applicants who have a higher chance to get $A$ and an identical chance to get $C$.

Alternatively, this scheme can be viewed as mimicking affirmative action. Suppose that the district identifies a target group of uninformed applicants who would acquire information if more seats at school $B$ are available to them. Being uninformed, this group is under-represented in $B$. The district accounts for the equilibrium response of this group and allocates more seats in $B$ to them. The group becomes informed, positively affecting other applicants.

This intervention would be harder to implement than the first one as a real-life policy because the district would need to know the distribution $\mathcal{F}(x)$ and the cost function $c(i)$ to anticipate the applicants' response correctly. However, many of the details are necessary only to ensure that the intervention benefits both informed and uninformed. A real-life implementation would probably have distributional concerns as at least one other objective; thus, the results of this example can be read that these policies, properly designed, may have desirable informational consequences.

I consider the same model as in Section \ref{sec:DAA}, except that the district allocates $q^0_A, q^0_B$ seats to uninformed and $q^1_A, q^1_B$ to informed applicants, at schools $A$ and $B$ respectively. Let $\gamma^\pti$ be the equilibrium fraction of informed applicants under these quotas and $r^0(A;\gamma^\pti),$ $r^0(B;\gamma^\pti),$ $r^1(A;\gamma^\pti)$ and $r^1(B;\gamma^\pti)$ be equilibrium cutoffs for uninformed and informed applicants respectively.\footnote{Note that there are two groups of applicants, informed and uninformed, which are treated differently. Hence, $\gamma^\pti$ -- the fraction of informed applicants -- also separates two distinct groups of applicants.} Anticipating the cutoffs, applicants decide whether to acquire information (and face cutoffs $r^1(A;\gamma^\pti)$ and $r^1(B;\gamma^\pti)$) or not (and face $r^0(A;\gamma^\pti)$ and $r^0(B;\gamma^\pti)$). The district selects seat allocation so that $r^1(B;\gamma^\pti) > r(B;\gamma^{Eq})$. Informed applicants value seats at school $B$ more than uninformed ones; thus, the value of information increases. Uninformed applicants are compensated for the loss of seats in $B$ by seats in $A$.

\begin{theorem}\label{th:RD_priority_design}
	Suppose that $p < \nicefrac{1}{3}$ and the cost function $c(i)$ is convex. There exist $\gamma^\pti \in (\gamma^{Eq},1)$, seat allocation $q_A^0, q_B^0, q_A^1, q_B^1$, and an equilibrium such that applicants $i \leq \gamma^\pti$ are informed, applicants $i > \gamma^\pti$ are not, and all applicants submit truthful preferences. In this equilibrium, applicants $i > \gamma^\pti$ have the same expected utility as under RSD and applicants $i \leq \gamma^\pti$ are better off.
\end{theorem}

Note that the theorem does not guarantee that $\gamma^\pti = \gamma^{SO}$. Indeed, if $\gamma^{Eq}$ is close to 1, then there may be too few uninformed applicants for the redistribution to achieve social optimum. This policy does not introduce a cost distortion, unlike the policy of revealing priorities ex-ante: the lowest-cost applicants acquire information.

\subsection{Tax-subsidy scheme}
	
	In this policy, learning is subsidized by imposing a flat tax on all applicants. Specifically, $\tau$ units of utility are collected from everyone. Then at least $\kappa > 0$ fraction of $\tau$ is used to subsidize learning of the target group of applicants with indices $i \in [\gamma^{Eq},\gamma^\tau]$, where $\gamma^\tau$ is a policy variable. The parameter $\kappa$ may capture the district's inability to perfectly identify the target group (e.g., the district may have to subsidize every informed applicant); wider economic distortions of taxation; or inefficiencies in the utility transfer. Note that $\kappa$ may be larger than one because the centralized provision of information may be more efficient than an individual search, for example, when information is provided in other languages for non-native speakers. For simplicity, I assume that $(1-\kappa)\tau$ is lost and does not enter anyone's utility; any other assumption would only strengthen my result.
	
	The applicants in the target group are given a subsidy $c(i) - c(\gamma^{Eq})$, so that their cost becomes $c(\gamma^{Eq})$. The tax needed for this scheme is
\begin{equation}\label{eq:tau}
	\tau (\gamma^\tau) = \frac{1}{\kappa} \int_{\gamma^{Eq}}^{\gamma^\tau} \left(c(x) - c(\gamma^{Eq})\right) dx,
\end{equation}	
and the utility of applicant $i \in [\gamma^{Eq},\gamma^\tau]$ in the tax-subsidy scheme describe above is
\begin{align*}
	U^\tau_i = U(e(i),\hat{R}_i|\gamma^\tau) - \tau(\gamma^\tau) - e(i) \min\{c(i),c(\gamma^{Eq})\}.
\end{align*}
Note that $U^0_i$ corresponds to the case where there is no subsidy and $\gamma^\tau = \gamma^{Eq}$. Note also that, once I allow for utility transfers, the normalization $u_A = 1$, $u_C = 0$ is no longer without loss of generality. 
 
\begin{theorem}\label{th:tax}
	Let $p < 1/3$. For each $\gamma > \gamma^{Eq}$, let $\tau(\gamma)$ be the total tax collected from all applicants, given by formula (\ref{eq:tau}). 
	Then for any $\kappa > 0$, there exists $\gamma^\tau > \gamma^{Eq}$ such that:
	\begin{enumerate}[(i)]
		\item every applicant $i \in [\gamma^{Eq},\gamma^\tau]$ is subsidized so that $c(i) = c(\gamma^{Eq})$ and
		\item every applicant is better off: for every $i \in [0,1]$, $U^\tau_i > U^0_i$.
	\end{enumerate}
\end{theorem}

The proof is based on two observations: (i) the worst-off applicant is uninformed and (ii) the utility of uninformed applicant increases with $\gamma$.

\section{Immediate Acceptance mechanism}\label{sec:IA}

In this section, I show that, when the costs are not ``too low,'' another popular assignment mechanism, Immediate Acceptance (IA), provides higher incentives to acquire information than does RSD. IA is the mechanism that has been widely used until recently, most prominently in Boston, but that has been criticized for its lack of strategy-proofness and for producing unfair (unstable) assignments \citep{abdulkadiroglu/sonmez:03,abdulkadiroglu/pathak/roth/sonmez:05,pathak2008leveling,ergin/sonmez:05}.\footnote{
		An extensive follow-up literature, starting with \citet{Abdulkadirouglu/Che/Yasuda-Boston_reconsidered:2011} and \citet{Miralles-For_Boston:2008}, argue that DA does not sufficiently account for cardinal preferences.} 

I focus exclusively on comparing informational incentives between two mechanisms and do not analyze welfare. The reason is that the problems with IA, well-documented in the literature, cannot be addressed in the environment of this paper. It should be mentioned that, for a range of parameters, IA does not Pareto dominate RSD: uninformed applicants are worse off in IA.

Throughout this section, I impose an additional assumption that cumulative distribution function $\mathcal{F}(x)$ is continuous on $[-\nicefrac{1}{2},\nicefrac{1}{2}]$.

As in RSD, each applicant decides (i) whether to acquire information ($e_i \in \{0,1\}$) and (ii) what ROL $R_i(\cdot)$ to submit ($R_i(\cdot) \in \{ABC,$ $ AC,$ $ BAC,$ $ BC, C\}$). As $C$ can accommodate all applicants, schools listed below $C$ are irrelevant and excluded. Denote applicant's strategy by $S_i = (e_i,R_i(\cdot))$. $R_i(\cdot)$ may depend on the acquired information about $B$. Applicants do not observe information decisions and ROLs of others. Each applicant is assigned a random priority $r_i$, unknown to applicants at the time they submit their ROLs.

The district collects applicants' ROLs and assigns applicants using IA algorithm, which, in the given environment, is described as follows. For applicant $i$,
\begin{itemize}
	\item If the quota in $i$'s top-ranked school is not exhausted by applicants who rank that school as their own top choice and whose priority is better than $r_i$, $i$ is assigned her top-ranked school; otherwise $i$ is rejected.
	\item If $i$ is rejected from her top-ranked school, then that school is removed from $R_i$; the quotas in all schools are reduced by the mass of applicants assigned to them in the current round; the process is repeated.
	\item The algorithm stops when all applicants are assigned.
\end{itemize}

The outcome of the IA algorithm can be expressed as a sequence of cutoffs $\rho^k_A(\gamma),$ $\rho^k_B(\gamma)$, corresponding to the $k$th round of the algorithm. The cutoffs are defined as follows: $\rho^k_X(\gamma) = \max\{r_i|i \text{ is accepted to }X\text{ in round }k\}$, where $X \in \{A,B\}$ (a notation slightly different from RSD's $r(A;\gamma), r(B;\gamma)$ is used for brevity). If no applicant is accepted to school $X$ in round $l$, $\rho^l_X(\gamma)=0$. For notational simplicity, I drop superscript for round one, so that $\rho_X(\gamma) = \rho^1_X(\gamma)$.

There always exists an equilibrium where schools $A$ and $B$ are full after round 1, which can informally be described as follows:
\begin{itemize}
	\item If $c(i)$ is ``high'': the fraction of informed applicants is $\gamma \in [0,\nicefrac{2}{3}]$; informed applicants submit truthful ROL;
	uninformed randomize between $ABC$ and $BAC$.
	\item If $c(i)$ is ``low'': $\gamma \in [\nicefrac{2}{3},1]$; there is $\bar \epsilon \in [0, \nicefrac{1}{2})$ such that informed applicants submit $ABC$ when $\epsilon_i < \bar \epsilon$ and $BAC$ otherwise; and all uninformed submit $ABC$.
\end{itemize}
The formal statement of the result is Lemma \ref{lemma_IA-ABfull} in Appendix~\ref{sec:appendix}. In the proof of Lemma \ref{lemma_IA-ABfull}, it is also established the cutoffs $\rho_A(\gamma)$ and $\rho_B(\gamma)$ do not change when $\gamma \in [0,\nicefrac{2}{3}]$. Thus, utility of uninformed applicants remains the same for $\gamma \in [0,\nicefrac{2}{3}]$. As that utility is the same under RSD and IA when $\gamma = 0$, and increases for RSD but not for IA, it means that uninformed applicants are worse off under IA if $\gamma^{IA} \in [0,\nicefrac{2}{3}]$ (see Corollary \ref{corr:RD_IA}).

The next theorem states that when the costs are not too high or too low, IA incentivizes learning better than RSD. The formal statement is in Appendix \ref{sec:appendix}.

\begin{theorem}\label{th:more_under_IA}
	Suppose that cost function $c(x)$ is such that $\gamma^{IA} > 0$ and $\gamma^{Eq} < \nicefrac{8}{9}$. Then $\gamma^{IA} > \gamma^{Eq}$.
\end{theorem}

The theorem applies to any equilibria of IA, not only the one described in Lemma \ref{lemma_IA-ABfull}; learning is higher in other equilibria, if they exist.

\begin{example}\label{ex-IA-IA-more-DA}
This example shows that if $\gamma^{Eq} > \nicefrac{8}{9}$, then $\gamma^{Eq} > \gamma^{IA}$ is possible.

	Let $c(x) = 3q/4$. Suppose that $\epsilon_i$ can take one of the four values: $-\nicefrac{3}{2}$ and $\nicefrac{3}{2}$ with probabilities $\nicefrac{1}{3}$ and $-.499, .499,$ with probabilities $\nicefrac{1}{6}$.\footnote{This probability distribution does not satisfy the assumption on $\mathcal{F}(x)$ maintained through the section, but the choice of $-.499$ and $.499$ ensures that the results are still valid.} Plugging in these values in the equations characterizing equilibria of RSD and IA, I obtain $\gamma^{Eq} = 1 > 0.97 \approx \gamma^{IA}$.
\end{example}

\section{Simulations}\label{sec:Simulations}

	My main model has three schools, with the value of the middle-ranked school being uncertain. This is a very special environment. This section presents simulations suggesting that learning may be insufficient for a wide range of parameters in environments with more than three schools.

	Let $\mathcal{S}$ be the set of schools, with $|\mathcal{S}| = N \in \{3, \dots, 6\}$ and identical quotas $\nicefrac{1}{N}$. There is a mass 1 of applicants. Applicant $i$'s utility from school $s_k$ is $u_i(s_k) = U(s_k) + \epsilon_i$, where $U(s_k)$ is $s_k$'s expected value, identical for all applicants, and $\epsilon_i \sim \mathcal{F}_{k}(x)$. The distribution $\mathcal{F}_{k}(x)$ is either normal or uniform, with standard deviations $(\sigma_{1}, \dots,\sigma_k,\dots, \sigma_{N})$. I assume that the applicant who pays the cost learns utilities from \emph{all} schools. The cost function is $c(i) = c \times i$.

	The simulations are conducted as follows. $U(s_k) = N-k$ for each $k$. $\sigma_{k} \in \{0,\dots,6\}$, for each $s_k$ independently. There are 200 cost parameters $c$, selected so that $\gamma^{Eq} \in \{0.005, 0.01, \dots, 1\}$. For each $\gamma^{Eq}$, and induced cost $c(i)$, I find $\gamma^{SO}$ and check whether $\gamma^{Eq} \geq \gamma^{SO}$. That is, there are 200 simulations for each profile $(\sigma_{1}, \dots, \sigma_{N})$, $N^7$ such profiles, and two types of distributions.
	
	I discard simulations where the assumption on school selectivity, maintained throughout the paper, is violated. I assume that if $s_k$ is above $s_j$ in the public ranking, then $r(s_k;\gamma) < r(s_j;\gamma)$ for any $\gamma \in [0,1]$. To see why this assumption may be violated, let schools $s_1,$ $s_2,$ and $s_3$ be ranked first, second and third in expectation and have identical quotas. Suppose that, in realization, $s_3$ is ranked above $s_1$ with probability $\nicefrac{1}{2}$ and $s_1$ is always ranked above $s_2$. When no one is informed, $r(s_1;0) < r(s_2;0) < r(s_3;0)$. However, when $\gamma = 1$, $s_1$ is top-ranked for half of the applicants and $s_3$ is for the other half. Then it must be the case that $r(s_1;1) = r(s_3;1) < r(s_2;1)$, violating the condition on school selectivity. Although the simulations do not allow the probability $\nicefrac{1}{2}$, the same effect is present when a variance of a low-ranked school is high and the variance of higher-ranked schools is low. I discard a simulation if there are two schools $s_k$ and $s_j$ such that $r(s_k;0) < r(s_j;0)$, but $r(s_k;1) > r(s_j;1)$.
	
	In Table \ref{tab:freq}, I report the fraction of simulations where the assumption of school selectivity is not violated and $\gamma^{SO} \leq \gamma^{Eq}$. That fraction is very small (far less than 1\%) and falling rapidly as the number of schools increases.
	
\begin{table}[h!]
	\centering \footnotesize
	\caption{Frequency of environments with $\gamma^{SO} \leq \gamma^{Eq}$}\label{tab:freq}
\begin{tabular}{ccccccccccc}
\toprule
           & \multicolumn{4}{c}{Normal}  &  & \multicolumn{4}{c}{Uniform} \\
					\cline{2-5}\cline{7-10}
Number of schools & 3    & 4    & 5     & 6     &  & 3     & 4    & 5    & 6     \\
Percentage with $\gamma^{SO}\leq\gamma^{Eq}$ & 0.09 & 0.02 & 0.007 & 0.001 &  & 0.26  & 0.09 & 0.03 & 0.009\\
\bottomrule
\end{tabular}
	
	\begin{tabnotes}
		Reported is the percentage of environments where $r(s_k;0) < r(s_j;0)$ implies $r(s_k;1) < r(s_j;1)$ and $\gamma^{SO} \leq \gamma^{Eq}$.
	\end{tabnotes}	
\end{table}%
	
One important question is in what kind of environments applicants do not under-acquire information. One way to answer this question is to find the environments where \emph{over-acquisition} of information is the largest. These environments are reported in Table \ref{tab:highdiff}. The highest fraction of applicants who over-acquire information is always observed in environments similar to the one studied in the theory section: the variance of the second-ranked school is high, while the variance of other schools is low.\footnote{My model can be readily extended to environments where there are $N$ schools, but only the second-ranked school is uncertain.}

\begin{table}[h!]
	\centering \footnotesize
	\caption{Environments with the largest $\gamma^{Eq} - \gamma^{SO}$}\label{tab:highdiff}
	\begin{tabular}{cccccccccccccccc}
		\toprule
	\multirow{2}{*}{$\substack{\text{No of}\\\text{schools}}$}&\multicolumn{7}{c}{Normal distribution}&\multicolumn{7}{c}{Uniform distribution}
	\\
	\cline{2-8}\cline{10-16}
	&{$\gamma^{Eq}-\gamma^{SO}$}& $\sigma_1$ & $\sigma_2$ & $\sigma_3$ & $\sigma_4$ & $\sigma_5$ & $\sigma_6$ &
	& {$\gamma^{Eq}-\gamma^{SO}$}& $\sigma_1$ & $\sigma_2$ & $\sigma_3$ & $\sigma_4$ & $\sigma_5$ & $\sigma_6$ 
	\\
	\midrule
	\multirow{2}{*}{3}&0.045 & 0 & 6 & 0 & -- & -- & -- & 
	&\multirow{2}{*}{0.075} & \multirow{2}{*}{0} & \multirow{2}{*}{6} & \multirow{2}{*}{1} & \multirow{2}{*}{--} & \multirow{2}{*}{--} & \multirow{2}{*}{--} 
	\\
  &0.045 & 0 & 6 & 1 & -- & -- & -- & 
	& & & & & & & 
	\\
	\midrule
	\multirow{2}{*}{4}&\multirow{2}{*}{0.045} & \multirow{2}{*}{0} & \multirow{2}{*}{6} & \multirow{2}{*}{0} & \multirow{2}{*}{0} & \multirow{2}{*}{--} & \multirow{2}{*}{--} &
	& 0.07  & 1 & 6 & 0 & 0 & -- & -- 
	\\
	&& & & & & & & 
	& 0.07  & 0 & 6 & 0 & 0 & -- & -- 
	\\
	\midrule
	\multirow{2}{*}{5}&\multirow{2}{*}{0.045} & \multirow{2}{*}{0} & \multirow{2}{*}{6} & \multirow{2}{*}{0} & \multirow{2}{*}{0} & \multirow{2}{*}{0} & \multirow{2}{*}{--} & 
	& 0.07  & 0 & 6 & 0 & 0 & 0 & -- 
	\\
	&& & & & & & & 
	& 0.07  & 1 & 6 & 0 & 0 & 0 & -- 
	\\
	\midrule
	\multirow{2}{*}{6}&\multirow{2}{*}{0.045} & \multirow{2}{*}{0} & \multirow{2}{*}{6} & \multirow{2}{*}{0} & \multirow{2}{*}{0} & \multirow{2}{*}{0} & \multirow{2}{*}{0} & 
	& 0.07  & 0 & 6 & 0 & 0 & 0 & 0
	\\
	&& & & & & & & 
	& 0.07  & 1 & 6 & 0 & 0 & 0 & 0 \\
	\bottomrule
  \end{tabular}%
	
	\begin{tabnotes}
		For each environment with $N$ school and either normal or uniform distribution, I find the tuple $(\sigma_{1}, \dots, \sigma_{N})$ where $\gamma^{Eq} - \gamma^{SO}$ is the largest and report the difference and the tuple in the table. $\gamma^{Eq}$ refers to the equilibrium fraction of informed applicants; $\gamma^{SO}$ is to the social optimum given $c(i) = c \times i$ corresponding to $\gamma^{Eq}$ and $\sigma_{k}$ is the standard deviation of the distribution of $\epsilon_i$ for school $s_k$. In all the environments, the largest $\gamma^{Eq} - \gamma^{SO}$ is reached when $\gamma^{Eq} = 1$.
	\end{tabnotes}	
\end{table}%

In Appendix \ref{sec:appendix4}, I provide an additional table that answers the following question: ``What are the low-variance environments where we observe $\gamma^{SO} \leq \gamma^{Eq}$?'' There is a larger variety of environments like that, but those with the second-ranked school being the only high-variance school still feature prominently.

\section{Discussion}\label{sec:Discussion}

	\subsection{School priorities}\label{sec:priorities}

	I have assumed that applicant's priorities are determined by a lottery only. In a typical school choice context, ``in-zone'' applicants are prioritized over the others. I argue below that the results and intuition carry over to this environment. As the ranking of applicants is no longer identical across schools, the usual strategy-proof mechanisms -- DA and Top Trading Cycle (TTC) -- are no longer equivalent to RSD. Below I separately consider information acquisition in DA and TTC.
	
	I start with DA. 
	Let $\pmb a, \pmb b, \pmb n$ 
	be groups of applicants in zones of schools $A$, $B$, and of neither school, respectively. That is, if $i \in \pmb g$ and $j \notin \pmb g$, where $\pmb g \in \{\pmb a, \pmb b\}$, then $i$ has a higher priority at $\pmb g$ than $j$. I assume that an applicant is never rejected from her zone school. The priorities of out-of-zone applicants are determined randomly, as in the main model; $r(A;\gamma)$ and $r(B;\gamma)$ refer to the cutoffs faced by out-of-zone applicants.
	
	As different groups of applicants face different cutoffs, they may make different information acquisition decisions. I denote the fractions of informed applicants in zones of $A$, $B$ and neither school by $\gamma^{\pmb a}$, $\gamma^{\pmb b}$ and $\gamma^{\pmb n}$; $\pmb\gamma = (\gamma^{\pmb a}, \gamma^{\pmb b}, \gamma^{\pmb n})$. For each of these groups, the equation $\Delta U^{\pmb g}(\pmb\gamma) = c(\gamma^{\pmb g}), \pmb g \in \{\pmb a, \pmb b, \pmb n\}$ determines the fractions of informed applicants.
	
	Social welfare is defined, as before, as the sum of utilities of all applicants:
	\begin{align*}
		SW&(\pmb\gamma) = \sum_{\pmb g \in \{\pmb a, \pmb b, \pmb n\}}\left[\gamma^{\pmb g} \Delta U^{\pmb g}(\pmb{\gamma})  + U^{\pmb g}(0,ABC|\pmb{\gamma}) - \int_0^{\gamma^{\pmb g}} c(i)di\right]
	\end{align*}
	
	It can be shown that applicants in group $\pmb b$ always under-acquire information and applicants in groups $\pmb a$ and $\pmb n$ under-acquire information if $|\pmb n| \geq |\pmb b|$, $p \leq 1/3$, and $\mathcal{I}$ is sufficiently small. The derivations are similar to those in Section \ref{sec:redesign_priority}, which discusses the policy of altering applicants' priorities in response to their information acquisition decisions.
	
	The analysis of TTC is more straightforward and similar to that of ex-ante known priorities because priorities in TTC provide a ``lower bound'' to applicants' scores. As in the analysis of DA, I assume that the mass of prioritized applicants is sufficiently small so that none of them are rejected from their priority schools. As school $A$ is the most popular, any applicant with the priority in $A$ can trade $A$ for any school; this applicant is equivalent to the ``high-priority'' applicant in Section \ref{sc:revealing}. If the gain $\mathcal{G}$ is not much higher than the cost, these applicants under-acquire information. An applicant with priority in $B$ can either get a score high enough for $A$ or if not, can trade $B$ for $C$ if she wants. This applicant is a ``combination'' of high- and mid-priority applicant form Section \ref{sc:revealing}. These applicants would under-acquire information under conditions milder than these in Section \ref{sec:SO} because they are less likely to cause displacement. Applicants with no priority at these two schools would have low incentives to learn and a low probability to be assigned to either $A$ or $B$.

\subsection{The comparison of RSD and IA when priorities are known ex-ante}

	One intervention I discuss in the paper is an ex-ante revelation of priorities (Section~\ref{sc:revealing}). Suppose that priorities are also ex-ante revealed under IA. In this case, the mechanisms become identical. Indeed, if applicant $i$ knows priority ex-ante, then, in equilibrium, $i$ knows which school she will be assigned to following the submission of her ROL: there is no uncertainty about priority and, given that there is a continuum of applicants, there is no uncertainty about aggregate preferences. The two mechanisms become identical and provide identical incentives.
	
\section{Conclusion}\label{sec:conclusion}

	Strategy-proof mechanisms have been widely advocated as replacements of manipulable mechanisms such as Immediate Acceptance, but their interactions with a larger environment require more research. This paper suggests that such mechanisms may provide low incentives to acquire information, and their adoption may need to be followed by policies that promote information acquisition. I discuss interventions that encourage learning, which could motivate a design of information policies suitable for real-life environments.

\setstretch{1}

\section{Acknowledgments}

{
I am indebted to the editor and anonymous referees whose questions and suggestions helped to improve the paper significantly. 
I am grateful to Ivan Balbuzanov, Suren Basov, Scott Kominers, and Bobby Pakzad-Hurson for suggestions and insightful discussions. 
I thank 
Estelle Cantillon, 
Yeon-Koo Che,
David Delacretaz, 
Guillaume Haeringer, 
Yinghua He, 
Fuhito Kojima, 
Jacob Leshno, 
Shengwu Li, 
Simon Loertscher, 
Matt Jones, 
Michael Ostrovsky, 
Marek Pycia, 
Alvin Roth, 
William Thompson, 
Utku Unver, 
Steven Williams, 
seminar participants at Buenos Aires, Deakin, Higher School of Economics, Hitotsubashi, Monash, Stanford, Tsukuba, Matching in Practice Workshop (Toulouse), SCW Conference (Boston), a Workshop at Victoria University of Wellington, International Workshop of Game Theory Society (Sao Paulo) for their comments. I gratefully acknowledge support from the Australian Research Council grant DP160101350 and a Faculty Research Grant at the University of Melbourne.
}

\newpage
\ifx\undefined\BySame
\newcommand{\BySame}{\leavevmode\rule[.5ex]{3em}{.5pt}\ }
\fi
\ifx\undefined\textsc
\newcommand{\textsc}[1]{{\sc #1}}
\newcommand{\emph}[1]{{\em #1\/}}
\let\tmpsmall\small
\renewcommand{\small}{\tmpsmall\sc}
\fi

\clearpage
\appendix

\pagenumbering{arabic}
\renewcommand*{\thepage}{A--\arabic{page}}

\setcounter{footnote}{0}
\renewcommand{\thefootnote}{A.\arabic{footnote}}

\setcounter{subsection}{0}
\renewcommand{\thesubsection}{A.\arabic{subsection}}

\setcounter{equation}{0}
\renewcommand{\theequation}{A.\arabic{equation}}
\setcounter{table}{0}
\renewcommand{\thetable}{A.\arabic{table}}
\setcounter{figure}{0}
\renewcommand{\thefigure}{A.\arabic{figure}}

\section*{Appendix to ``Assignment mechanisms: common
preferences and information acquisition''}

\subsection{Omitted proofs}\label{sec:appendix}

\subsubsection{Proofs from Section \ref{sec:DAA}}

First, we need to establish that truthful reporting is a weakly dominated strategy. As $\gamma$ is fixed below, I suppress it in the notation. 
Consider an applicant who takes cutoffs $r(A)$ and $r(B)$ as given. Suppose that applicant $i$ picks strategy $s_i = (e_i, XYZ)$, where $XYZ$ is an arbitrary permutation of $\{A,B,C\}$. Then the expected utility of applicant $i$ is
\begin{align}
    Eu_i&(e_i,XYZ) 
    =\mbox{Pr}(r_i \leq r(X)) Eu_i(X) + \mbox{Prob}(r(X) < r_i \leq r(Y))Eu_i(Y) \nonumber\\
    &+ \mbox{Pr}(\max\{r(X),r(Y)\}<r_i\leq r(Z))Eu_i(Z) - e_i c(i) \nonumber\\
    =& r(X) Eu_i(X) + \max\{r(Y)-r(X),0\} Eu_i(Y) \qquad\left(\because r_i \sim U[0,1]\right)\nonumber\\
    &+ \max\{r(Z)-\max\{r(X),r(Y)\},0\}Eu_i(Z) - e_i c(i)
\end{align}

\begin{claim}\label{claim:truthful}
    Let $Eu_i(X)>Eu_i(Y)>Eu_i(Z)$. Then
    \begin{equation}
        Eu_i(e_i, XYZ) \geq Eu_i(e_i, R),\label{eq:two_ROLs}
    \end{equation}
    where $R$ is an arbitrary permutation of the set $\{X,Y,Z\}$. 
		
		Furthermore, if $r(A)>0$, $r(B)>0$, $r(C)=1$ and equation (\ref{eq:two_ROLs}) holds with equality, then $R$ generate the same probability distribution over school assignments as $XYZ$.
\end{claim}

\begin{proof}[Proof of Claim \ref{claim:truthful}]\label{proof:truthful}

First, note that all the permutations of schools $\{X, Y, Z\}$ form a group with respect to the operation of composition of permutations. I call this operation ``group multiplication'' and denote it by $*$. A well-known property of this group (indeed, of any $S_n$) is that any permutation can be written as a product of transpositions (a transposition is a permutation that exchanges the places of just two elements, leaving all other elements intact).\footnote{For the general theory of permutation groups see, for example, \cite{Dixon-Mortimer_Permutation_Groups}.} Now, group $S_3$ has three transpositions: $(1,2), (2,3)$ and $(1,3)$; here $(i,k)$ denotes the transposition that exchanges places of elements $i$ and $k$. However, only two of transpositions are independent, because $(1,3) = (1,2)*(2,3)$. Therefore, it is sufficient to prove Claim \ref{claim:truthful} only for two independent transpositions, $(1,2)$ and $(2,3)$.

(i) If $Eu_i(X) > Eu_i(Y)$ then 
\vspace{-20pt}

\begin{align}
    Eu_i&(e_i,XYZ)  - Eu_i(e_i, YXZ)\nonumber\\
		=&r(X) Eu_i(X) + \max\{r(Y)-r(X),0\} Eu_i(Y) - r(Y) Eu_i(Y)\nonumber\\
		 -& \max\{r(X)-r(Y),0\} Eu_i(X) = \min\{r(X),r(Y)\}(Eu_i(X) - Eu_i(Y)) \geq 0,\label{app_eq:two_ROLs_i}
\end{align}
with strict inequality if $\min\{r(X),r(Y)\}>0$. Note that this result does not depend on the utility from $Z$.

(ii) If $Eu_i(Y) > Eu_i(Z)$ then
\vspace{-20pt}

\begin{align}
		Eu_i&(e_i, XYZ) - Eu_i(e_i, XZY)\nonumber\\
		=&\max\{r(Y)-r(X),0\} Eu_i(Y) + \max\{r(Z) - \max\{r(X),r(Y)\},0\} Eu_i(Z)\nonumber\\ 
		-& \max\{r(Z)-r(X),0\} Eu_i(Z) - \max\{r(Y) - \max\{r(X),r(Z)\},0\} Eu_i(Y)\nonumber\\
		=&\max\{\min\{r(Y), r(Z)\} - r(X), 0\}(Eu_i(Y) - Eu_i(Z)) \geq 0,\nonumber
\end{align}
with strict inequality if $\min\{r(Y), r(Z)\} > r(X)$. This result is independent of the utility from $X$.

Thus, the first part of Claim \ref{claim:truthful} is established.

To prove the second part of Claim \ref{claim:truthful}, first note that $\min\{r(X),r(Y)\}>0$, hence inequality in (\ref{app_eq:two_ROLs_i}) is strict. That is, $X$ must be listed as the top school and I only need to compare $XYZ$ and $XZY$. If $XYZ$ generates a different probability distribution than $XZY$, then $u_i(e_i, XYZ) > u_i(e_i, XZY)$, which contradicts the statement of the theorem.\footnote{Note that if $r(A) < r(B) < r(C)$, then an applicant whose true preferences are $BAC$ gets the same probability distribution over possible assignments when she submits $BAC$, her true preferences, and $BCA$. This is because if she is rejected from $B$, then $r_i > r(B) > r(A)$ and she will also be rejected from school $A$.} 
\end{proof}

Given Claim \ref{claim:truthful}, I will assume that every applicant reports either the truthful ROL -- the ROL that corresponds to the applicant's expected utilities -- or an ROL that generates the same probability distribution over school assignments as the truthful ROL.

Next, I establish that, for any $\gamma$, if applicant $i$ is rejected from school $B$, then $A$ is not feasible for $i$.

\begin{claim}\label{rArBclaim}
	For any $\gamma \in [0,1]$, $r(A;\gamma) \leq r(B;\gamma)$; $r(A;\gamma) = r(B;\gamma)$ only if $\gamma = 1$ and $p = 1/2$.
\end{claim}

\begin{proof}[Proof of Claim \ref{rArBclaim}]\label{rArBproof}
    The proof is by contradiction. Suppose that $r(A;\gamma) \geq r(B;\gamma)$; thus, anyone rejected from $A$ is assigned to $C$. The mass $\gamma p$ lists school $B$ as the top choice. 
		
		Consider first the case where $\gamma p \geq q$. Then equating supply (LHS of equations (\ref{eq:appendix_r_Ar_B1}) and (\ref{eq:appendix_r_Ar_B2})) and demand (RHS of equations (\ref{eq:appendix_r_Ar_B1}) and (\ref{eq:appendix_r_Ar_B2})) gives us
    \begin{align}
    	q = & r(B;\gamma) \gamma p \label{eq:appendix_r_Ar_B1}\\
    	q = & r(A;\gamma) \left[(1-\gamma) + \gamma (1 - p)\right]+ (r(A;\gamma) - r(B;\gamma)) \gamma p
			= r(A;\gamma) - q\label{eq:appendix_r_Ar_B2}
    \end{align}

    In turn, this implies the following conditions on the cut-offs:
    \begin{align*}
    	r(A;\gamma) = 2q,\quad r(B;\gamma) =  \frac{q}{\gamma p}
    \end{align*}

    Recall that $p \leq 1/2$ and $\gamma \leq 1$, thus $\frac{1}{\gamma p} \geq 2$, hence $r(A;\gamma) \leq (B;\gamma)$, with strict inequality if $\gamma < 1$ or $p < 1/2$ -- a contradiction to our initial supposition.

    Consider now the case $\gamma p < q$. In this case every applicant to $B$ is accepted, including an applicant with $r_i = 1$, so $r(B;\gamma) = 1$. Yet, as we assumed $r(A;\gamma) \geq (B;\gamma)$, it means $r(A;\gamma)=1$ as well. Thus, schools $A$ and $B$ accept all applicants, but the total quota of these two schools is $2q \leq 2/3 < 1$, a contradiction.
\end{proof}

Claim \ref{rArBclaim} allows me to write supply-demand equations as in (\ref{eq:general-rA}) and (\ref{eq:r_B_general_v}).

\begin{proof}[Proof of Theorem \ref{th:Gamma_nonempty}]
	Define mapping $\phi: [0,1] \mapsto [0,1]$ as $\gamma = i^*(\Delta U(\gamma)) \equiv \phi(\gamma)$. This mapping is continuous and into itself. Therefore, by Brower's fixed point theorem, it has a fixed point.
\end{proof}

\begin{proof}[Proof of Theorem \ref{th:eqSD}]
	I first show that the strategy profile $\{(\hat e_i, \hat R_i)\}_{i \in N}$ corresponding to $\gamma^{Eq} \in \Gamma$ as defined by (\ref{th_eq:RD_e_i})-(\ref{th_eq:RD_R_i}) is a Nash equilibrium. 
	
	Suppose applicant $i$ deviates to $(e_i',R_i') \neq (\hat e_i, \hat R_i)$. By Claim \ref{claim:truthful}, $Eu_i(e_i', \hat R_i) \geq Eu_i(e_i', R_i')$. Hence, I can focus on a deviation to $(e_i', \hat R_i)$. Note that, after the deviation, the fraction of informed applicants is still $\gamma^{Eq}$. Since $\gamma^{Eq} \in \Gamma$, $\Delta U(\gamma^{Eq}) = c(\gamma^{Eq})$. Thus, if $i \leq \gamma^{Eq}$, then $Eu_i(1, \hat R_i) \geq Eu_i(0, \hat R_i)$ and if $i < \gamma^{Eq}$, then $Eu_i(0, \hat R_i) > Eu_i(1, \hat R_i)$. A deviation to $(e_i', R_i')$ does not increase the utility of $i$: $(\hat e_i, \hat R_i)$ is a Nash equilibrium. 

	Next, I show that if $\{(e_i', R_i')\}_{i \in N}$ is a Nash equilibrium, then the fraction of informed applicants $\gamma'$ induced by $\{e_i'\}_{i \in N}$ is in $\Gamma$ and the assignment is the same as under strategy $\{(e_i', \hat R_i)\}_{i \in N}$ for all applicants but possibly $i$.
	
	Suppose that $R_i' \neq \hat R_i$ for some $i$. Then $Eu_i((e_i', R_i'),(e_j', R_j')_{j \neq i}) = Eu_i((e_i', \hat R_i),(e_j', R_j')_{j \neq i})$, as otherwise $i$ would deviate to $(e_i', \hat R_i)$. By Claim \ref{claim:truthful}, the probability distribution over $i$'s assignment is the same under these two strategy profiles; $\{(e_i', \hat R_i),(e_j', R_j')_{j \neq i}\}$ is also a Nash equilibrium. Repeating the argument for each $i$ for whom $R_i' \neq \hat R_i$, we conclude that $\{(e_i', \hat R_i)\}_{i \in N}$ is a Nash equilibrium that induces the same probability distribution over applicants' assignments as $\{(e_i', R_i')\}_{i \in N}$.

	Suppose that $\gamma' \notin \Gamma$; then either $c(\gamma') < \Delta U(\gamma')$ or $c(\gamma') > \Delta U(\gamma')$.
	
	First, consider the case $c(\gamma') < \Delta U(\gamma')$. As $c(x)$ is a continuous function, there exists $i' > \gamma'$ such that $c(i') < \Delta U(\gamma')$. Since $c(i) < \Delta U(\gamma')$ for all $i \leq i'$, then any $i$ with $e_i = 0$ has a profitable deviation to $(1, \hat R_i)$. As $\{(e_i', \hat R_i)\}_{i \in N}$ is an equilibrium, we conclude that all $i \leq i'$ must be informed. Thus, the fraction of informed applicants is at least $i'$. As $i' > \gamma'$, it contradicts the initial assumption that $\gamma'$ fraction of applicants is informed. 
	
	Similarly, if $c(\gamma') > \Delta U(\gamma')$, there is $i' < \gamma'$ such that $c(i') > \Delta U(\gamma')$. In equilibrium, applicants $i > i'$ must be uninformed; thus, at least $(1 - i') > (1 - \gamma')$ fraction of applicants is uninformed, which contradicts the assumption that $\gamma'$ fraction is informed.
	
	Therefore, if $\{(e_i',\hat R_i)\}_{i \in N}$ is a Nash equilibrium, then $\gamma' \in \Gamma$. 
	
	By a similar argument, applicants with index $i < \gamma'$ do and with $i > \gamma'$ do not acquire information; thus $e_i' = \hat e_i$ for all $i \neq \gamma'$. Thus, for all applicants $i \neq \gamma'$, $\{(e_i',\hat R_i)\}_{i \in N}$ is described by (\ref{th_eq:RD_e_i})-(\ref{th_eq:RD_R_i}) and leads to the same information decision and assignment as $\{(e_i', R_i')\}_{i \in N}$.	
\end{proof}

\begin{proof}[Proof of Theorem \ref{th:Pareto_ranking}]
	
	There are three groups of applicants: $[0,\gamma_L]$ who are informed in both equilibria; $(\gamma_L,\gamma_H]$, who are informed in $\gamma_H$, but remain uninformed in $\gamma_L$ equilibrium; and $(\gamma_H,1]$, who are uninformed in both equilibria.

    The first and the last groups are better off under $\gamma_H$ compared to $\gamma_L$ because their utility positively depend on $r(A;\gamma)$ and $r(B;\gamma)$, which are, in turn, increasing: see formulae (\ref{eq:U(0|gamma)}) and (\ref{eq:U(1|gamma)}). Thus, I need to evaluate the effect on the group $(\gamma_L, \gamma_H]$; these applicants are informed in only one of the equilibria.

	Since for any $i < \gamma_H$, $U(1,\hat{R}|\gamma_H) - c(i) \geq U(1,\hat{R}|\gamma_H) - c(\gamma_H)$ and applicant $\gamma_H$ chooses to acquire information, then
    \begin{align*}
        U(1,\hat{R}|\gamma_H) - c(i) \geq
        U(1,\hat{R}|\gamma_H) - c(\gamma_H) \geq U(0,ABC|\gamma_H) > U (0,ABC|\gamma_L).
    \end{align*}
    Thus, every applicant $i \in (\gamma_L,\gamma_H]$ is better off in the equilibrium associated with $\gamma_H$.
\end{proof}

\begin{proof}[Proof of Theorem \ref{th:SO-more-than-DA}]
	Recall that $\gamma^{Eq}$ solves $\Delta U(\gamma^{Eq}) = c(\gamma^{Eq})$. Thus, the derivative of the social welfare function is
	\[
		\frac{\partial}{\partial \gamma} SW(\gamma^{Eq}) = \gamma^{Eq} \frac{\partial}{\partial \gamma}\Delta U(\gamma^{Eq}) + \frac{\partial}{\partial \gamma} U(0,ABC|\gamma^{Eq}).
	\]
	Recall that $U(0,ABC|\gamma)$ and $\Delta U(\gamma)$ increase with $r(A;\gamma), r(B;\gamma)$ (see equations \ref{eq:U(0|gamma)} and \ref{eq:U(1|gamma)}). Recall also that $\frac{\partial}{\partial \gamma}r(A;\gamma) > 0$ and $\frac{\partial}{\partial \gamma}r(B;\gamma) \geq 0$ for $\gamma \in [0,\nicefrac{1}{3p}]$ (see equations \ref{rA} and \ref{rB}). Thus, $\frac{\partial}{\partial \gamma} SW(\gamma^{Eq}) > 0$ when $\gamma^{Eq} \in [0,\nicefrac{1}{3p}]$.
\end{proof}
Note that the last line does not imply that $\gamma^{SO} = 1$ because when $\gamma > \gamma^{Eq}$, the first term of (\ref{eq:dSW}) is negative.

\begin{proof}[Proof of Theorem \ref{th:Pareto_improvement_revelation}]

    When applicants know their priorities, they also know which schools are feasible for them in equilibrium; their decisions will be based on that. Let $\pmb a$ be a group for whom $A$ is feasible and $\pmb b$ be a group for whom $A$ is not feasible but $B$ is. Denote the fraction of informed applicants in $\pmb a$ by $\gamma_{\pmb a}$ and in $\pmb b$ by $\gamma_{\pmb b}$. Note that applicants not belonging to these two groups are assigned to $C$ regardless of what ROL they submit, so all of them are uninformed (except, possibly, a zero-cost applicant). By fixed point arguments similar to those in Section \ref{sec:DAA}, equilibrium exists with $\gamma_{\pmb a}, \gamma_{\pmb b}$ determined by equations $\Delta U_{\pmb a}(\gamma_{\pmb a}, \gamma_{\pmb b}) = c(\gamma_{\pmb a})$ and $\Delta U_{\pmb b}(\gamma_{\pmb a}, \gamma_{\pmb b}) = c(\gamma_{\pmb b})$, where $\Delta U_{\pmb g}$ is the net gain of learning for $\pmb g \in \{\pmb a, \pmb b\}$, given by
	\begin{align}
		\Delta U_{\pmb g}(\gamma_{\pmb a}, \gamma_{\pmb b}) = p \mathcal{G}\label{eq:reveal_eq_cond}
	\end{align} 
	
	The formula is the same for both groups, implying $\gamma_{\pmb a} = \gamma_{\pmb b} \coloneqq \gamma^*$. Note that the formula is similar to (\ref{eq:diffutil}) except for $r(B;\gamma)$. To see why, consider group $\pmb a$. Applicants in $\pmb a$ are guaranteed a seat in $A$. If they learn that $B$ is better than $A$, they are assigned to $B$ with probability 1; their expected gain is $p \left(E[u_i(B)|u_i(B)>1] - 1\right) = p \mathcal{G}$. Similarly, applicants in $\pmb b$ are guaranteed a seat in $B$ and would only go to $C$ instead of $B$ if they learn that $C$ is better for them than $B$. The gain is $p \left(1/2 - E[u_i(B)|u_i(B)<0]\right) = p \mathcal{G}$.

	Cutoffs for schools $A$ and $B$ (or groups $\pmb a$ and $\pmb b$) are $r_{\pmb a}(\gamma^*,\gamma^*)$ and $r_{\pmb b}(\gamma^*,\gamma^*)$ can be calculated by equating supply of and demand for seats in schools $A$ and $B$:
	\begin{align}
		&r_{\pmb a}(\gamma^*,\gamma^*) = q\frac{1}{1-\gamma^* p}, \label{eq:revealing_a}
		\\
		&r_{\pmb b}(\gamma^*,\gamma^*) = q\frac{2 - 2\gamma_{\pmb a} p - \gamma_{\pmb b} p}{(1-\gamma_{\pmb a} p)(1-\gamma_{\pmb b} p)}= q\frac{2 - 3\gamma^* p}{(1-\gamma^* p)^2}, \label{eq:revealing_b}
	\end{align}
	Equations (\ref{eq:revealing_a}), (\ref{eq:revealing_b}) are identical to cutoffs $r(A;\gamma^{Eq})$ and $r(B;\gamma^{Eq})$ (equations~\ref{rA}, \ref{rB}), except for different $\gamma^*$ and $\gamma^{Eq}$. Recall that $\gamma^{Eq}$ is determined by equation $r(B;\gamma^{Eq}) p \mathcal{G} = c(i)$; as $r(B;\gamma) < 1$ and $c(i)$ is strictly increasing, it implies $\gamma^* > \gamma^{Eq}$, $r_{\pmb a}(\gamma^*,\gamma^*) > r(A;\gamma^{Eq})$ and $r_{\pmb b}(\gamma^*,\gamma^*) > r(B;\gamma^{Eq})$. Then, for any applicant, the assignment under $r_{\pmb a}(\gamma^*,\gamma^*), r_{\pmb b}(\gamma^*,\gamma^*)$ is weakly better than under $r(A;\gamma^{Eq}), r(B;\gamma^{Eq})$.
	
	Next, I show that there is a positive mass of applicants who are better off. Consider arbitrary applicant $i$ with priority $r_i \in \big(r(A;\gamma^{Eq}),r_{\pmb a}(\gamma^*,\gamma^*)\big)$, who is uninformed when priorities are unknown and submits $ABC$. Suppose $i$ remains uninformed when priorities are known and submits $ABC$. Then $i$ is assigned to $B$ under unknown priorities and $A$ under known priorities. Hence, if $i$ remains uninformed, she is better off; she will only increase her utility if she decides to learn. Hence, there is a positive mass of applicants who are better off.
\end{proof}

	To state Theorem \ref{th:revealing_ranks_underacquisition} formally, I need to define social welfare and the cost cutoff. I define social welfare as the sum of all utilities, taking into account that uninformed applicants with priority $r_i \leq r_{\pmb a}$ have utility 1 and with priority $r_{\pmb a} < r_i \leq r_{\pmb b}$ have utility 1/2; and applicants with $r_i > r_{\pmb b}$ have utility 0: 
	\begin{align}
		SW(\gamma_{\pmb a}, \gamma_{\pmb b}) 
		\begin{aligned}[t]
		&= r_{\pmb a}(\gamma_{\pmb a}, \gamma_{\pmb b})\left(1 + \gamma_{\pmb a} p \mathcal{G} - \int_0^{\gamma_{\pmb a}}c(x)dx\right)\\
		&+ (r_{\pmb b} - r_{\pmb a})(\gamma_{\pmb a}, \gamma_{\pmb b})\left(\frac{1}{2} + \gamma_{\pmb b} p \mathcal{G} - \int_0^{\gamma_{\pmb b}}c(x)dx\right).\label{eq:revealing_SW}
		\end{aligned}
	\end{align}
	The cost cutoff is defined as
	\[C = 2\gamma^* p \left(1 + \int^{\gamma^*}_0(p\mathcal{G} - c(x))dx\right),\]
	where $\gamma^*$ is defined above. 
	The formal statement of Theorem \ref{th:revealing_ranks_underacquisition} is below.
	\setcounter{theorem}{5}
	\begin{theorem}
		Under the policy of revealing priorities before information acquisition,
		\begin{enumerate}
			\item Among applicants with $r_i < r_{\pmb a}$, too few are informed if $C < 1$ and too many if $C>1$;
			\item Too few applicants with priorities $r_{\pmb a} < r_i \leq r_{\pmb b}$ are informed;
			\item Applicants with priorities $r_i > r_{\pmb b}$ are uninformed and it is socially optimal.
		\end{enumerate}
	\end{theorem}
\begin{proof}[Proof of Theorem \ref{th:revealing_ranks_underacquisition}]

	Differentiating $SW(\gamma_{\pmb a}, \gamma_{\pmb b})$ defined in (\ref{eq:revealing_SW}) with respect to $\gamma_{\pmb a}$ and $\gamma_{\pmb b}$, and taking into account that $r_{\pmb a}$ does not depend on $\gamma_{\pmb b}$, I get
	\begin{align*}
		\frac{\partial SW(\gamma_{\pmb a}, \gamma_{\pmb b})}{\partial \gamma_{\pmb a}} 
		&= \frac{1}{2}\frac{\partial}{\partial \gamma_{\pmb a}}(r_{\pmb a} + r_{\pmb b})(\gamma_{\pmb a}, \gamma_{\pmb b}) + \frac{\partial r_{\pmb a}(\gamma_{\pmb a}, \gamma_{\pmb b})}{\partial \gamma_{\pmb a}}\left((\gamma_{\pmb a} - \gamma_{\pmb b})p\mathcal{G} + \int_{\gamma_{\pmb a}}^{\gamma_{\pmb b}}c(x)dx\right) 
			\\&
			+ r_{\pmb a}(\gamma_{\pmb a}, \gamma_{\pmb b})\left(p\mathcal{G} - c(\gamma_{\pmb a})\right) 
			+ \frac{\partial r_{\pmb b}(\gamma_{\pmb a}, \gamma_{\pmb b})}{\partial \gamma_{\pmb a}} \int_0^{\gamma_{\pmb b}}\left(p\mathcal{G} - c(x)\right)dx
			\\
			\frac{\partial SW(\gamma_{\pmb a}, \gamma_{\pmb b})}{\partial \gamma_{\pmb b}} &=  \frac{\partial r_{\pmb b}(\gamma_{\pmb a}, \gamma_{\pmb b})}{\partial \gamma_{\pmb b}}\left(\frac{1}{2} + \int_0^{\gamma_{\pmb b}}\left(p\mathcal{G} - c(x)\right)dx\right) +  (r_{\pmb b} - r_{\pmb a})(\gamma_{\pmb a}, \gamma_{\pmb b})\left(p\mathcal{G} - c(\gamma_{\pmb b})\right)
	\end{align*}
	Recall that, in equilibrium, $\gamma_{\pmb a} = \gamma_{\pmb b} = \gamma^*$ and $p\mathcal{G} = c(\gamma^*)$; thus, in equilibrium, the derivatives collapse to:
	\begin{align}
		\frac{\partial SW(\gamma^*,\gamma^*)}{\partial \gamma_{\pmb a}} &= \frac{1}{2}\frac{\partial}{\partial \gamma_{\pmb a}}(r_{\pmb a} + r_{\pmb b})(\gamma^*,\gamma^*) + \frac{\partial r_{\pmb b}(\gamma^*,\gamma^*)}{\partial \gamma_{\pmb a}} \int_0^{\gamma^*}\left(p\mathcal{G} - c(x)\right)dx \label{eq:revelation_SW_a}
		\\
		\frac{\partial SW(\gamma^*,\gamma^*)}{\partial \gamma_{\pmb b}} &=  \frac{\partial r_{\pmb b}(\gamma^*,\gamma^*)}{\partial \gamma_{\pmb b}}\left(\frac{1}{2} + \int_0^{\gamma^*}\left(p\mathcal{G} - c(x)\right)dx\right).\label{eq:revelation_SW_b}
	\end{align}
	As $p\mathcal{G} > c(i)$ for any $i < \gamma^*$, 
	\[\int_0^{\gamma^*}\left(p\mathcal{G} - c(x)\right)dx > 0.\]
	Thus, the signs of the derivatives depend on $\frac{\partial}{\partial \gamma_{\pmb a}}\left(r_{\pmb a} + r_{\pmb b}\right)(\gamma^*,\gamma^*)$, $\frac{\partial r_{\pmb b}(\gamma^*,\gamma^*)}{\partial \gamma_{\pmb a}}$ and $\frac{\partial r_{\pmb b}(\gamma^*,\gamma^*)}{\partial \gamma_{\pmb b}}$, which are given below:
	\begin{align*}
		&\frac{\partial}{\partial \gamma_{\pmb a}}\left(r_{\pmb a} + r_{\pmb b}\right)(\gamma^*,\gamma^*) = \frac{qp(1 - 2\gamma_{\pmb b} p)}{(1-\gamma_{\pmb a} p)^2(1-\gamma_{\pmb b} p)} = \frac{q p(1 - 2\gamma^*p)}{(1-\gamma^*p)^3} > 0
		\\
		&\frac{\partial r_{\pmb b}(\gamma^*,\gamma^*)}{\partial \gamma_{\pmb a}} = -\frac{q\gamma_{\pmb b}p^2}{(1-\gamma_{\pmb a}p)^2(1-\gamma_{\pmb b}p)}  = -\frac{q\gamma^*p^2}{(1-\gamma^*p)^3} < 0
		\\
		&\frac{\partial r_{\pmb b}(\gamma^*,\gamma^*)}{\partial \gamma_{\pmb b}} = \frac{qp(1 - 2\gamma_{\pmb a}p)}{(1-\gamma_{\pmb a}p)(1-\gamma_{\pmb b}p)^2}  = \frac{qp(1 - 2\gamma^*p)}{(1-\gamma^*p)^3} > 0
	\end{align*}
  Plugging in the derivatives into equation (\ref{eq:revelation_SW_a}), I obtain the condition that $\frac{\partial SW(\gamma^*,\gamma^*)}{\partial \gamma_{\pmb a}} > 0$ if $C < 1$, while the derivative $\frac{\partial SW(\gamma^*,\gamma^*)}{\partial \gamma_{\pmb b}} > 0$ for any $\gamma^*$. 
\end{proof}

\begin{proof}[Proof of Theorem \ref{th:RD_priority_design}]

The proof is based on the following claim.

\begin{claim}\label{claim:priority}
	Suppose $\gamma^{Eq} \in (0,1)$. Fix the actions of the applicants: 
	\begin{itemize}
		\item $i \in [0,\gamma^{Eq}]$ acquire information, 
		\item $i \in (\gamma^{Eq},1]$ do not acquire information, and
		\item all applicants report their preferences truthfully. 
	\end{itemize}
	Let $\gamma^p \in [\gamma^{Eq},1)$. Suppose that applicants $i \in (\gamma^p, 1]$ are assigned to schools $A$ and $B$ with quotas $(q_A^0, q_B^0)$ and applicants $i \in [0,\gamma^p]$ are assigned to schools $A$ and $B$ with quotas $(q - q_A^0, q - q_B^0)$, where
	\begin{gather}\label{eq:priorities_quotas}
	(q_A^0, q_B^0) = 
	\begin{cases}
		\left(\frac{1}{2}(1-\gamma^p)(r(A;\gamma^{Eq}) + r(B;\gamma^{Eq})), 0\right) & \text{ if } (1-\gamma^p)(r(A;\gamma^{Eq})+r(B;\gamma^{Eq})) \leq 2q\\
		\left(q, (1 - \gamma^p)(r(A;\gamma^{Eq}) + r(B;\gamma^{Eq})) - 2q\right) & \text{ if } (1-\gamma^p)(r(A;\gamma^{Eq})+r(B;\gamma^{Eq})) > 2q.
	\end{cases}
	\end{gather}
	Then the cutoffs of schools $A$ and $B$ for the two groups of applicants, $r^0(A;\gamma^{Eq}),$ $r^0(A;\gamma^{Eq}),$ $r^1(A;\gamma^{Eq}),$ $r^1(B;\gamma^{Eq})$, are such that
	\begin{align}
		r(A;\gamma^{Eq}) &\geq r^1(A;\gamma^{Eq})\label{eq:priority_cutoffs_A}\\
		r^1(B;\gamma^{Eq}) &\geq r(B;\gamma^{Eq})\label{eq:priority_cutoffs_B}\\
		r^0(A;\gamma^{Eq}) &= (r(A;\gamma^{Eq}) + r(B;\gamma^{Eq}))/2\label{eq:priority_cutoffs_AB0}\\
		r^1(A;\gamma^{Eq}) + r^1(B;\gamma^{Eq}) &> r(A;\gamma^{Eq}) + r(B;\gamma^{Eq})\label{eq:priority_cutoffs_AB1}
	\end{align}
\end{claim}

\begin{proof}

	Consider first the case where $(1-\gamma^p)(r(A;\gamma^{Eq})+r(B;\gamma^{Eq})) \leq 2q$.
	
	Given that $(1 - \gamma^p)$ applicants are assigned to $q^0_A$ seats in school $A$, $r^0(A;\gamma^{Eq})$ satisfies equation $r^0(A;\gamma^{Eq})(1 - \gamma^p) = q_A^0 = (1 - \gamma^p)(r(A;\gamma^{Eq}) + r(B;\gamma^{Eq}))/2 $. Thus, $r^0(A;\gamma^{Eq}) = (r(A;\gamma^{Eq})+r(B;\gamma^{Eq}))/2$, establishing (\ref{eq:priority_cutoffs_AB0}).
	
	Cutoff $r^1(A;\gamma^{Eq})$ equates supply of and demand for seats in $A$ from informed applicants whose mass is $\gamma^p$:
	\begin{align}
		q - (1-\gamma^p) (r(A;\gamma^{Eq})+r(B;\gamma^{Eq}))/2 = r^1(A;\gamma^{Eq}) (\gamma^p - \gamma^{Eq} p),\label{eq:priority_A}
	\end{align}
	which can be rewritten, using that, from (\ref{eq:general-rA}), $q = r(A;\gamma^{Eq})(1-\gamma^{Eq}p)$, as 
	\begin{align*}
		r(A;\gamma^{Eq})(1-\gamma^{Eq}p) - (1-\gamma^p) (r(A;\gamma^{Eq})+r(B;\gamma^{Eq}))/2 = r^1(A;\gamma^{Eq}) (\gamma^p - \gamma^{Eq}p)
	\end{align*}		
	As $r(B;\gamma^{Eq}) > r(A;\gamma^{Eq})$ (see Claim \ref{rArBclaim}), 
	\begin{align*}
		(1-\gamma^{Eq}p)r(A;\gamma^{Eq}) - (1-\gamma^p) r(A;\gamma^{Eq}) = (\gamma^p -\gamma^{Eq}p)r(A;\gamma^{Eq}) > r^1(A;\gamma^{Eq}) (\gamma^p - \gamma^{Eq}p),
	\end{align*}
	and $r(A;\gamma^{Eq}) > r^1(A;\gamma^{Eq})$ follows.
	
	Similarly, cutoff $r^1(B;\gamma^{Eq})$ is determined by equation
	\begin{align*}
		q &= r^1(B;\gamma^{Eq}) \gamma^{Eq} p + (r^1(B;\gamma^{Eq}) - r^1(A;\gamma^{Eq}))(\gamma^p - 2p\gamma^{Eq}).
	\end{align*}	

	Combining equations above, I obtain
	\begin{gather}
		((r^1(B;\gamma^{Eq}) + r^1(A;\gamma^{Eq})) - (r(B;\gamma^{Eq})+r(A;\gamma^{Eq})))(\gamma^p - \gamma^{Eq}p) = (r(A;\gamma^{Eq}) - r^1(A;\gamma^{Eq})) \gamma^{Eq} p.\label{eq:uninformed_priority}
	\end{gather}
	Recall that $r(A;\gamma^{Eq}) > r^1(A;\gamma^{Eq})$. From (\ref{eq:uninformed_priority}), $r^1(A;\gamma^{Eq}) + r^1(B;\gamma^{Eq}) > r(A;\gamma^{Eq}) + r(B;\gamma^{Eq})$ and, from this inequality and (\ref{eq:priority_cutoffs_A}), $r^1(B;\gamma^{Eq}) > r(B;\gamma^{Eq})$.
	
	Consider now the case $(1-\gamma^p)(r(A;\gamma^{Eq})+r(B;\gamma^{Eq})) > 2q$. Recall that no places in school $A$ are available to applicants with $i \in (\gamma^p, 1]$, hence all of them go either to $B$ or $C$. Cutoff $r^1(B;\gamma^{Eq})$ is determined by the equation
	\begin{align}
		q - q_B^0 = \gamma^p (1-p) r^1(B;\gamma^{Eq}).\label{eq:school_B_priority_case2}
	\end{align}
	which can be combined with
	\begin{equation}
		3q = r(A;\gamma^{Eq}) \gamma^{Eq} p + (r(B;\gamma^{Eq})+r(A;\gamma^{Eq}))(1-\gamma^{Eq}p), \label{eq:3q_DA}
	\end{equation}
	which itself follows from (\ref{eq:general-rA}) and (\ref{eq:r_B_general_v}),
	%
	to obtain
	\begin{align*}
		r(A;\gamma^{Eq}) \gamma^p + p (\gamma^p - \gamma^{Eq}) r^1(B;\gamma^{Eq}) = (\gamma^p - \gamma^{Eq}p) (r^1(B;\gamma^{Eq}) - r(B;\gamma^{Eq})).
	\end{align*}
	As $\gamma^p \geq \gamma^{Eq}$, the LHS of the equation is positive; $\gamma^p - \gamma^{Eq}p$ is also positive, so $r^1(B;\gamma^{Eq}) > r(B;\gamma^{Eq})$.
	
	Finally, using equation (\ref{eq:school_B_priority_case2}), I obtain
	\begin{gather*}
		\gamma^p (r(B;\gamma^{Eq})+r(A;\gamma^{Eq})) \leq \gamma^p r^1(B;\gamma^{Eq}) + (r(B;\gamma^{Eq}) - r^1(B;\gamma^{Eq})) \gamma^{Eq} p < \gamma^p r^1(B;\gamma^{Eq}),
	\end{gather*}
	thus (\ref{eq:priority_cutoffs_AB1}) follows.
\end{proof}

	We are now in a position to prove Theorem \ref{th:RD_priority_design}.

	Note that Claim \ref{claim:truthful} applies: there is no profitable deviation from submitting the truthful ROL $\hat{R}_i$.

	Fix $x \in (\gamma^{Eq}, 1)$ and the choice of quotas $(q_A^0,q_B^0,q_A^1,q_B^1)$ corresponding to $x$, given by equation (\ref{eq:priorities_quotas}) with $\gamma^p = x$. Suppose also that mass $x$ of applicants are informed. Note that $x$ may not correspond to any equilibrium. The resulting cutoffs are $r^0(A;x), r^0(B;x), r^1(A;x), r^1(B,x)$.

	Consider function \[G(x) = U^1((1,\hat{R})|x) - c(x) - U((0,ABC)|x),\] 
	where $U^1((1,\hat{R})|x) = U((1,\hat{R})|r^1(A;x), r^1(B;x))$ is the expected utility of targeted informed applicants and $ U((0,ABC)|x) =  U((0,ABC)|r(A;x), r(B;x))$ is expected utility of uninformed applicants before the intervention
	(recall that the index in utility function $U$ is suppressed because the expression is the same for all applicants); this function is defined for $x \in [\gamma^{Eq},1]$. 
	
	Next, I show that there exists $s$ such that $G(x) = 0$. 

	First, note that
	\begin{align}
		G(\gamma^{Eq})
		&> U^1((1,\hat{R})|\gamma^{Eq}) - c(\gamma^{Eq}) - U^1((0,ABC)|\gamma^{Eq})\label{eq:G_greater_0_step1}
		\\
		&> U((1,\hat{R})|\gamma^{Eq}) - U((0,ABC)|\gamma^{Eq}) - c(\gamma^{Eq}) = 0,\label{eq:G_greater_0}
	\end{align} 
	where (\ref{eq:G_greater_0_step1}) follows from (\ref{eq:priority_cutoffs_AB1}); inequality between (\ref{eq:G_greater_0_step1}) and (\ref{eq:G_greater_0}) follows from equation (\ref{eq:diffutil}) and (\ref{eq:priority_cutoffs_B}); and (\ref{eq:G_greater_0}) from $\gamma^{Eq}$ being equilibrium.
	
	Consider now $\gamma=1$. In that case, $q_A^1 = q_B^1 = q$ and $r^1(A;1) = r(A;1),$ $r^1(B;1) = r(B;1)$. Thus $U^1((1,\hat{R})|1) = U((1,\hat{R})|1)$, substituting it in $G(x)$, we get
	\begin{equation} 
		G(1) = U((1,\hat{R})|1) - U((0,ABC)|1) - c(1) < 0.\label{eq:G_smaller_0}
	\end{equation} 
	where the last inequality follows from $\gamma = 1$ not being an equilibrium in the environment with unknown priorities.
	
	As $G(x)$ is a linear combination of continuous functions, it is continuous. As $G(\gamma^{Eq}) > 0 > G(1)$, there exists $x$ such that $G(x) = 0$. I denote it by $\gamma^\pti$. 
	
	Finally, note that for any $x \in [\gamma^{Eq},1)$, $U((0,ABC)|x) = U^0((0,ABC)|x),$ hence \[U^1((1,R)|\gamma^\pti) - c(\gamma^\pti) - U^0((0,ABC)|\gamma^\pti) = G(\gamma^\pti) = 0.\] Thus, applicant $i = \gamma^\pti$ is indifferent to learning. Hence, there is an equilibrium where $i \leq \gamma^\pti$ are informed and $i > \gamma^\pti$ are uninformed. 
	
	The utility of applicants $i \geq \gamma^\pti$ is the same in RSD with original quotas and in RSD with $q_A^0, q_B^0$. It follows directly from (\ref{eq:priority_cutoffs_AB0}) for applicants $i > \gamma^\pti$; for $i = \gamma^\pti$, it follows from the fact that $i$ is indifferent to acquiring information. 
	
	Thus, what remains to be shown is that $i < \gamma^\pti$ are better off. First, applicants $i \in [\gamma^{Eq}, \gamma^\pti)$, who are uninformed in RSD with $q_A = q_B = q$ but choose to acquire information under the seat re-allocation scheme, are weakly better off. Indeed, they can always deviate to acquiring no information and get the same utility as in the RSD with unknown priorities; given that they choose to acquire information in equilibrium means that this deviation is not profitable. 
Applicants with $i \leq \gamma^{Eq}$ are informed in both regimes. By Claim \ref{claim:priority} \emph{if} only $\gamma^{Eq}$ fraction were informed, then $r^1(B;\gamma^{Eq}) > r(B;\gamma^{Eq})$ and $r^1(A;\gamma^{Eq}) + r^1(B;\gamma^{Eq}) > r(A;\gamma^{Eq}) + r(B;\gamma^{Eq})$; hence these applicants are better off with quotas $(q_A^1, q_B^1)$ even if the fraction of informed applicants is the same. As the equilibrium fraction of informed applicants with quotas $(q_A^1,q_B^1)$ is higher than $\gamma^{Eq}$, and since $r^1(B;x)$ and $r^1(A;x) + r^1(B;x)$ increase with $x$, informed applicants are better off with quotas $(q_A^1, q_B^1)$.
\end{proof}

\begin{proof}[Proof of Theorem \ref{th:tax}]

	Consider the following maximization problem:
	\[
		\max_{\gamma \geq \gamma^{Eq}} U(0,ABC|\gamma) - \tau(\gamma)
	\]
	Note $\tau(\gamma^{Eq}) = 0$ and $U(0,ABC|\gamma) - \tau(\gamma)$ is increasing near $\gamma^{Eq}$. Pick some $\gamma^\tau$ such that 
	\[U(0,ABC|\gamma^\tau) - \tau > U(0,ABC|\gamma^{Eq}).\]
	
	The inequality above implies that uninformed applicants $i \in (\gamma^\tau,1]$ are better off. 
	
	Note that
	\begin{align}
		U^\tau_i=&U(0,ABC|\gamma^\tau) + \Delta U(\gamma^\tau) - \tau(\gamma^\tau) \geq U(0,ABC|\gamma^\tau) + \Delta U(\gamma^{Eq}) - \tau(\gamma^\tau) \nonumber 
		\\&> U(0,ABC|\gamma^{Eq}) + \Delta U(\gamma^{Eq}) = U^0_i.\label{eq:tax_informed}
	\end{align}
As $\Delta U(\gamma^{\tau}) > \Delta U(\gamma^{Eq}) \geq c(i)$, applicants in $[0,\gamma^{Eq}]$ acquire information under tax and (\ref{eq:tax_informed}) shows that they are better off.

	Finally, consider applicants $i \in [\gamma^{Eq},\gamma^\tau]$. Each of these applicants face the same cost, $c(\gamma^{Eq})$ and, for each of them, $\Delta U(\gamma^\tau) > \Delta U(\gamma^{Eq}) = c(\gamma^{Eq})$. Thus, they are informed. Then equation (\ref{eq:tax_informed}) applies to these applicants and they are also better off.
\end{proof}

To prove Theorem \ref{th:more_under_IA}, I first find the lowest-information equilibrium of IA.

\begin{lemma}\label{lemma_IA-ABfull}
    There exists a Bayes-Nash equilibrium $(e_i,R_i(\cdot))_{i \in N}$ of IA game where both schools $A$ and $B$ are full after round one. The equilibria are characterized by three values: the fraction of informed applicants $\gamma  \in [0,1]$, the threshold $\bar{\epsilon} \in [0, \nicefrac{1}{2})$, and the probability $\alpha \in [\nicefrac{2}{3},1]$. The strategies defined as follows:
    \begin{itemize}
      \item Every applicant $i < \gamma$ is informed ($e_i=1$); every applicant $i > \gamma$ is not informed ($e_i=0$); and applicant $i = \gamma$ may be either.
      \item Every informed applicant $i$ submits $R_i=BAC$ or $R_i = BC$ if $\epsilon_i>\bar{\epsilon}$ and submits $R_i=ABC$ or $R_i=AC$ if $\epsilon_i<\bar{\epsilon}$. An applicant $i$ with $\epsilon_i = \bar{\epsilon}$ submits $R_i \in \{BAC, BC, ABC, AC\}$.
      \item Mass $\alpha$ of uninformed applicants submits $ABC$ or $AC$ and mass $(1-\alpha)$ submits $BAC$ or $BC$.
    \end{itemize}

	The values $\gamma, \bar{\epsilon}$ and $\alpha$ are determined as follows.

	There are four non-overlapping conditions on costs: (Ia) $\mathcal{T}_1(0) < c(0)$; (Ib) $c(0) \leq \mathcal{T}_1(0) < c(2/3)$; (IIa) $c(2/3) \leq \mathcal{T}_1(0)$ and $c(1) < \mathcal{T}_2(\bar{\epsilon}^*)$; (IIb) $c(1) \geq \mathcal{T}_2(\bar{\epsilon}^*)$, where 
	\begin{align*}
		\mathcal{T}_1(z) = 3q \int^{\infty}_{z} x d\mathcal{F}(x),
	\end{align*}
	\[
		\mathcal{T}_2(z) = q\left[\frac{1}{\nicefrac{1}{2}+z} + 1\right]\int_{z}^{\infty} (x-z)d\mathcal{F}(x),
	\]
	$\bar{\epsilon}^*$ is the solution to equation
	\begin{equation}
		\mathcal{T}_3(\bar{\epsilon}^*) = 1, \label{eq:IA_cases}
	\end{equation}
	where
	\begin{equation*}
		\mathcal{T}_3(z) = \left[\frac{1}{\nicefrac{1}{2}+z} + 1\right] (1 - \mathcal{F}(\bar{\epsilon})). 
	\end{equation*}

	In region (Ia), $\gamma = 0$, $\alpha = 2/3$ and $\bar{\epsilon}$ is undetermined because all applicants are uninformed.

	In region (Ib), $\gamma$ is determined by equation $c(\gamma) = \mathcal{T}_1(0)$, $\alpha$ is determined by $(1-\alpha)(1-\gamma) = \nicefrac{1}{3} - \nicefrac{\gamma}{2}$ and $\bar{\epsilon} = 0$. The value of $\alpha$ changes from 2/3 to 1 as $\gamma$ changes from 0 to 2/3.

	In region (IIa), $\gamma$ and $\bar{\epsilon}$ are determined by equations 
	\begin{align}
		&\mathcal{T}_1(\bar{\epsilon}) = c(\gamma)\label{NE-IA-case1-eq2}\\
		&\mathcal{T}_3(\bar{\epsilon}) = 1/\gamma \label{NE-IA-case1-eq1}
	\end{align}
	and $\alpha = 1$.

	In region (IIb), $\gamma = 1$, $\alpha = 1$ and $\bar{\epsilon}$ are determined by equation (\ref{eq:IA_cases}).
	
	There are no other equilibria where both schools $A$ and $B$ are full after round 1.
\end{lemma}

\begin{proof}[Proof of Lemma \ref{lemma_IA-ABfull}]

	To prove Lemma \ref{lemma_IA-ABfull}, I first establish two claims.

	\begin{claim}\label{claim-IA-info}
	    Consider equilibrium $s^*=(e_i,R_i(\cdot))_{i \in N}$. If $e_j = 1$, then for any $i < j$, $e_i = 1$. If $e_j = 0$, then for any $i > j$, $e_i = 0$.
	\end{claim}

	\begin{proof}[Proof of Claim \ref{claim-IA-info}]
		Suppose that applicant $j$ acquires information. Since $s^*$ is a Nash equilibrium, then $U(1,R_j(\cdot)) - c(j) \geq U(0,R')$ for any $R'$. Since $c(j) > c(i)$ for any $i < j$, then $U(1,R_j(\cdot)) - c(i) > U(0,R')$ for any $R'$. Thus, $(0, R')$ cannot be a Nash equilibrium strategy of applicant $i$.
		
		The proof for the second part of the statement is identical.
	\end{proof}

	\begin{claim}\label{claim-IA-C}
		Ranking $C$ as the top school is strictly dominated. Formally, the following strategies are strictly dominated:
		\begin{itemize}
			\item $(1, R_i(x))$ such that	the set $\{x \in \mathbb{R}|R_i(x)\in\{C,CA,CB,CAB,CBA\}\}$ has a positive measure; and
			\item $(0, R_i)$ with $R_i\in\{C,CA,CB,CAB,CBA\}$
		\end{itemize}
	\end{claim} 

\begin{proof}[Proof of Claim \ref{claim-IA-C}]
	If $C$ is top-ranked in $R_i$, then applicant $i$ is assigned to $C$. If $A$ is ranked as the top school and $r_i \leq q$, then $i$ is guaranteed to be assigned to $A$. Consider a strategy $(e_i,R'_i)$ such that $R'_i(x) = R_i(x)$ for $x$ where $C$ is not top-ranked in $R_i(x)$ and $R'_i(x) = AC$ otherwise. Then the assignment is identical except in the cases when $i$ is assigned to $C$ under $R_i(x)$ and to $A$ under $R'_i(x)$. There is a positive measure of these cases, hence $U(e_i,R'_i(x)) > U(e_i,R_i(x))$.
\end{proof}
	
	I am now in a position to prove Lemma \ref{lemma_IA-ABfull}. I distinguish the following cases:
	(a) when all uninformed applicants submit $AC$; 
	(b) when uninformed applicants are indifferent between submitting $AC$ and $BC$; and 
	(c) when all uninformed applicants submit $BC$.

Claim \ref{claim-IA-info} establishes that there is a cutoff $\gamma \in [0,1]$ that determines information acquisition choice. Claim \ref{claim-IA-C} establishes that I need to consider strategies $\{ABC,AC,BAC,BC\}$ only. As I am looking for an equilibrium where schools $A$ and $B$ are full after the first round, applicant's assignment will not change if she submits $ABC$ or $AC$ and $BAC$ or $BC$. Thus, it is sufficient to consider $AC$ and $BC$ only. 

	Recall that if applicant $i$ lists $AC$ and has priority $r_i \leq \rho_A$, $i$ is assigned to $A$; if $i$ lists $BC$ and has priority $r_i \leq \rho_B$, $i$ is assigned to $B$; and $i$ is assigned to $C$ in all other cases. Thus, the expected utility of applicant $i$ when she submits ROLs $AC$ and $BC$ can be written as:

    \begin{equation}\label{eq-IA-Case1-utils}
        \begin{split}
        &U(e_i,AC) = \rho_A \times 1\\
        &U(e_i,BC) = \rho_B (\nicefrac{1}{2}+\epsilon)
        \end{split}
    \end{equation}

    For given $\rho_A, \rho_B$, define
    \begin{equation}\label{eq-IA-baru}
			\bar\epsilon = \frac{\rho_A}{\rho_B} - \frac{1}{2}.
    \end{equation}
    As $\rho_B>0$, $\bar{\epsilon}$ is well-defined. In equilibrium, applicant $i$ with $\epsilon_i>\bar{\epsilon}$ submits $BC$ and $i$ with $\epsilon_i < \bar{\epsilon}$ submits $AC$.

    Note that (a) if $\bar{\epsilon}>0$, all uninformed applicants (for whom $Eu_i(B) = \nicefrac{1}{2}$) submit $AC$; (b) if $\bar{\epsilon}=0$, uninformed applicants are indifferent; and (c) if $\bar{\epsilon}<0$, all uninformed applicants submit $BC$.

    \textit{Case (a)}: $\bar{\epsilon}>0$.

    If $\gamma(1-\mathcal{F}(\bar{\epsilon})) \geq q$, cutoffs $\rho_A$ and $\rho_B$ are determined by the following supply-demand equations:
	\begin{align}
		q &= 
		\rho_A \left(1-\gamma(1-\mathcal{F}(\bar{\epsilon}))\right)\label{eq-IA-case1a-rhoA}\\
		q &= \rho_B\gamma(1-\mathcal{F}(\bar{\epsilon}))\label{eq-IA-case1a-rhoB}
	\end{align}
	If $\gamma(1-\mathcal{F}(\bar{\epsilon})) < q$, then $\rho_B = 1$. I will rule out this case later.

	Combining (\ref{eq-IA-baru}), (\ref{eq-IA-case1a-rhoA}) and (\ref{eq-IA-case1a-rhoB}), I obtain (\ref{NE-IA-case1-eq1}):
    \[
    	\left[\frac{1}{\nicefrac{1}{2}+\bar{\epsilon}} + 1\right] (1 - \mathcal{F}(\bar{\epsilon})) = \frac{1}{\gamma}.
    \]
%
	Note that at $\bar{\epsilon} = 0$, $\gamma = 2/3$. As $\bar{\epsilon}$ increases, the LHS of equation (\ref{NE-IA-case1-eq1}) monotonically decrease. Thus, $\gamma$ is uniquely determined for every value of $\bar{\epsilon}$ and $\bar{\epsilon}(\gamma)$ is a monotonically increasing function of $\gamma$. At $\bar{\epsilon}=\nicefrac{1}{2}$, the LHS becomes $2(1-\mathcal{F}(\nicefrac{1}{2}))<1$, hence $\gamma > 1$. Thus, for any value of $\gamma \in (\nicefrac{2}{3},1]$, there exists a unique solution $\bar{\epsilon}(\gamma) \in (0,\nicefrac{1}{2})$.

	Next, I find applicant $i$ who is indifferent to learning. Recall that an uninformed applicant submits $AC$ in equilibrium; her expected utility is $U(0,AC) = \rho_A$ (equation \ref{eq-IA-Case1-utils}). If $i$ is informed, then, with probability $\mathcal{F}(\bar{\epsilon})$, $\epsilon_i < \bar{\epsilon}$ and $i$ submits $R_i = AC$; otherwise, $R_i = BC$. Thus, indifference condition can be written as
	\begin{align}
		U(1, &R_i(\cdot)) - U(0,AC) = (1-\mathcal{F}(\bar{\epsilon}))\rho_B E[u_i(B)|\epsilon_i > \bar{\epsilon}] + \mathcal{F}(\bar{\epsilon})\rho_A - \rho_A \nonumber
		\\
		&= (1-\mathcal{F}(\bar{\epsilon}))\rho_B (E[\epsilon_i|\epsilon_i>\bar{\epsilon}] - \bar{\epsilon}) \quad (\because \rho_A = (\nicefrac{1}{2}+\bar{\epsilon}) \rho_B)\nonumber 		\\
		&= q \left[\frac{1}{\nicefrac{1}{2}+\bar{\epsilon}} + 1\right]\int_{\bar{\epsilon}}^{\infty} (x-\bar{\epsilon})d\mathcal{F}(x) = c(i), \label{eq:IA_case_a}
	\end{align}
	In equation (\ref{eq:IA_case_a}), the LHS decreases and RHS increases with $\gamma$. Therefore, it has a solution $\gamma \in (2/3,1]$ only if
	\begin{align}
		c(2/3) \leq 3q\int_{0}^{\infty} x d\mathcal{F}(x) \label{eq:IA_case_a_cond1}
	\end{align}
	and if solution exists, it is unique. If equation (\ref{eq:IA_case_a_cond1}) holds, but either $\gamma(\bar{\epsilon}) > 1$ or solution does not exist, then $\gamma = 1$ and $\bar{\epsilon}$ is determined by equation
  \[
		\left[\frac{1}{\nicefrac{1}{2}+\bar{\epsilon}} + 1\right] (1 - \mathcal{F}(\bar{\epsilon})) = 1.
  \]
	This case applies when
	\[
		q\left[\frac{2}{1+2\bar{\epsilon}} + 1\right]\int_{\bar{\epsilon}}^{\infty} (x-\bar{\epsilon})d\mathcal{F}(x) \geq c(1).
	\]
	All applicants are informed; these with $\epsilon_i > \bar{\epsilon}$ submit $BC$ and these with $\epsilon_i < \bar{\epsilon}$ submit $AC$.
	
		I return now to the case where $\gamma(1-\mathcal{F}(\bar{\epsilon})) < q$ and $\rho_B = 1$. Recall that in Case (a) all uninformed applicants submit $AC$, so $U(0,AC) = \rho_A \geq {\rho_B}/{2} = \nicefrac{1}{2} = U(0,BC)$. That is, $\rho_A \geq \nicefrac{1}{2}$. At the same time, using equation (\ref{eq-IA-case1a-rhoA}) and conditions $\gamma(1-\mathcal{F}(\bar{\epsilon})) < q$ and $q \leq 1/3$, I obtain
		\begin{align*}
			\rho_A = \frac{q}{1 - \gamma(1 - \mathcal{F}(\bar{\epsilon}))} < \frac{q}{1-q} \leq \frac{1}{2},
		\end{align*}
		which is inconsistent with $\rho_A \geq 1/2$.
				
    \textit{Case (b)}: $\bar{\epsilon}=0$. Uninformed applicants are indifferent between $AC$ and $BC$. I assume that $\alpha$ fraction of uninformed applicants submit $AC$ and $(1-\alpha)$ fraction apply to $BC$. As $\bar{\epsilon} = 0$, one-half of all informed applicants submit $BC$ and one-half submit $AC$. Thus, the cutoffs $\rho_A$ and $\rho_B$ are given by:
    \begin{align*}
        &q = \rho_A(\gamma/2+(1-\gamma)\alpha),\\
        &q = \rho_B(\gamma/2+(1-\gamma)(1-\alpha))
    \end{align*}

    Given that uninformed applicants are indifferent between submitting $AC$ and $BC$, hence $U(0,AC) = U(0,BC)$, the following condition must hold:
    \begin{equation}\label{eq-IA-Case1-indiff-cond2}
        \rho_A = \rho_B\times \frac{1}{2},
    \end{equation}
    which implies, using the formulae for $\rho_A$ and $\rho_B$,
    \begin{align}
        3(\gamma/2 + (1-\gamma)(1-\alpha)) &= 1.\label{eq-IA-Case1b-alpha}
    \end{align}
    It then follows that $\rho_B = 3q$ and $\rho_A = 1.5q$. Furthermore, it must be that $\alpha \in [0,1]$ and $\gamma \in [0,1]$. Hence, equation (\ref{eq-IA-Case1b-alpha}) imposes the following conditions: $\gamma \in [0,2/3]$ and $\alpha \in [2/3,1]$.

		Applicant $i$ is indifferent to learning when the following equality holds:
    \begin{align*}
        &\frac{1}{2}\rho_B E[u_i(B)|u_i(B)>\nicefrac{1}{2}] + \frac{1}{2}\rho_A - c(i) = \rho_A.
    \end{align*}
    Using expression for $\rho_B$, equation (\ref{eq-IA-Case1-indiff-cond2}) and taking into account that $i = \gamma$, I obtain
    \begin{align*}
        \frac{3}{2}q \left(E\left[u_i(B)|u_i(B)>\frac{1}{2}\right]-\frac{1}{2}\right) = 3q \int_{0}^{\infty} x d\mathcal{F}(x) = c(\gamma).
    \end{align*}
    As the LHS of this equation is a constant and the RHS is monotonically increasing with $\gamma$, the solution with $\gamma \in [0,2/3]$ exists if only if 
		\begin{align*}
			c(0) \leq 3q \int_{0}^{\infty} x d\mathcal{F}(x) \leq c(2/3),
		\end{align*}
		and it is unique.

    If $3q \int_{0}^{\infty} x d\mathcal{F}(x) < c(0)$, then $\gamma=0$: no one acquires information. In that case, equation (\ref{eq-IA-Case1b-alpha}) becomes $3(1-\alpha)=1$, hence $\alpha = \nicefrac{2}{3}$. If $3q \int_{0}^{\infty} x d\mathcal{F}(x) > c(1)$, then implied $\gamma$ is equal to one, which leads to Case (a).
		
		\textit{Case (c)}: $\bar{\epsilon} < 0$. In this case, uninformed applicants submit $BC$; only informed applicants submit $AC$ and, given that $\bar{\epsilon} < 0$, less than half of them do so. Thus, $\rho_A > 2q$, as if $2q/\gamma < 1$, then $\rho_A  = 2q/\gamma \geq 2q$ and if $2q/\gamma \geq 1$ then $\rho_A = 1$. 
		
		The cutoff at school $B$ is determined by equation $q = \rho_B(\nicefrac{\gamma}{2} + (1-\gamma)) = \rho_B(1-\nicefrac{\gamma}{2})$. Hence $\rho_B = \frac{q}{1-\nicefrac{\gamma}{2}} \leq 2q$. Thus, $\rho_A > \rho_B$: anyone who is accepted to school $B$ would be accepted to school $A$, if applied. School $A$ is more valuable for both uninformed applicants and for informed applicants whose $\epsilon_i \in (\bar{\epsilon},\nicefrac{1}{2})$. Those applicants submit $BC$ in Case (c), but have a profitable deviation to $AC$. There is no equilibrium in this case. 
\end{proof}

\begin{proof}[Proof of Theorem \ref{th:more_under_IA}]

Before proving Theorem \ref{th:more_under_IA}, I establish two claims that allow me to consider only the equilibrium described in Lemma \ref{lemma_IA-ABfull}.

\begin{claim}\label{claim:IA_unfilled_B}
	If there is an equilibrium $(e_i,R_i(\cdot))$ of IA game such that school $B$ has unfilled seats in round 2, then the fraction of informed applicants, $\gamma^{IA2}$, is larger than the fraction of informed applicants in RSD game, $\gamma^{Eq}$.
\end{claim}

\begin{proof}[Proof of Claim \ref{claim:IA_unfilled_B}]
  As $B$ has unfilled seats in round 2, any applicant to $B$ in round 1 is accepted, so $\rho_B=1$. In round 2, applicants with $r_i \leq \rho_{B,2}$ are accepted. The first-round cutoff for $A$ is $\rho_A$ and, as $A$ is full after round 1, $\rho_{A,2}=0$. Note that only the applicants who are rejected from $A$ apply to $B$ in round two; then, for any such applicant $i$, $r_i > \rho_A$. As a positive mass of applicants is accepted to $B$ in round 2, $\rho_{B,2} > \rho_A$. The last inequality also implies that, for an uninformed applicant, $ABC$ yields a higher utility than $AC$, as she has a chance to gain admission to school $B$ if unsuccessful at school $A$.
	
	Let $\bar{\epsilon}$ be a solution to equation
	\begin{align}
		\rho_A + (\rho_{B,2} - \rho_A) (\bar{\epsilon}+\nicefrac{1}{2}) 
		= (\bar{\epsilon} + \nicefrac{1}{2}) \label{eq:IA_another_equil1}
	\end{align}
	The LHS is a utility of applicant $i$ whose $\epsilon_i = \bar{\epsilon}$ and who submits $ABC$; the RHS is $i$'s utility when she submits $BC$. Thus, any applicant with $\epsilon_i > \bar{\epsilon}$ submits $BC$ and any applicant with $\epsilon_i < \bar{\epsilon}$ submits $ABC$.
	Equation (\ref{eq:IA_another_equil1}) can be re-written as
	\begin{align*}
		\rho_{B,2} - \rho_A = 1 - \frac{\rho_A}{\bar{\epsilon}+\nicefrac{1}{2}}
	\end{align*}

	The difference in utilities between informed and uninformed applicants -- which depends on $\gamma$ through $\rho_A, \rho_{B,2}$ and $\bar{\epsilon}$ -- is
	\begin{gather*}
		\Delta U^{IA2} (\gamma) = \int_{\bar{\epsilon}}^{\infty} \left(x +\frac{1}{2}\right) d\mathcal{F}(x) - (\rho_{B,2}-\rho_A) \left(\int_{-\infty}^{-\frac{1}{2}} + \int_{\bar{\epsilon}}^{\infty}\right) \left(x+\frac{1}{2}\right) d\mathcal{F}(x) - \rho_A \int_{\bar{\epsilon}}^{\infty} d\mathcal{F}(x)
		\\\geq \int_{\frac{1}{2}}^{\infty} \left(x - \frac{1}{2}\right) d\mathcal{F}(x)	
	\end{gather*}
	
	Recall that $\Delta U (\gamma)  \leq 2.25q \int_{\nicefrac{1}{2}}^{\infty} (x - \nicefrac{1}{2}) d\mathcal{F}(x)$. As I assume that $q \leq \frac{1}{3}$, then $\Delta U(\gamma) < \Delta U^{IA2} (\gamma)$ for any $\gamma \in [0,1]$. Hence, $\gamma^{Eq} < \gamma^{IA2}$.
\end{proof}

\begin{claim}\label{claim:IA_no_unfilled_A}
    There is no equilibrium under IA where $A$ has unfilled seats after round 1.
\end{claim}

\begin{proof}[Proof of Claim \ref{claim:IA_no_unfilled_A}]
	Suppose not. Since school $A$ has unfilled seats after round 1, any applicant who prefers $A$ to $B$ would submit ROL that lists $A$ as the top choice. That is, $(1-\gamma)$ mass of uninformed applicants and $\gamma(1-p)$ mass of informed applicants will apply to $A$ in round 1. Thus, the total mass $(1-\gamma)+\gamma(1-p) = 1 - \gamma p \geq \nicefrac{1}{2}$ applies to $A$, while the quota at $A$ is at most $\nicefrac{1}{3}$. This is a contradiction to $A$ being unfilled after round 1.
\end{proof}

	I am now in a position to prove Theorem \ref{th:more_under_IA}.
	
	Recall that $\gamma^{Eq}$ solves the problem
	\begin{align}
			&c(\gamma^{Eq}) = q \frac{2 - 3\gamma^{Eq} p}{(1 - \gamma^{Eq} p)^2} \int^{\infty}_{\frac{1}{2}} \left(x - \frac{1}{2}\right) d\mathcal{F}(x)\label{eq:SD_for_IA_vs_SD}
	\end{align}
	
	Suppose that $\gamma^{Eq} < 8/9$ and that $\gamma^{IA} \leq \gamma^{Eq}$.
	
	If $\gamma^{IA} < 2/3$, then $\gamma^{IA}$ solves 
	\begin{align}
		c(\gamma^{IA}) = 3q\int_{0}^{\infty} x d\mathcal{F}(x).\label{eq:IA_caseIb_for_SD_vs_IA}
	\end{align}
	If $2/3 \leq \gamma^{IA} < 8/9$, then $\gamma^{IA}$ solves
	\begin{align}
		&c(\gamma^{IA}) = \frac{q}{\gamma^{IA}(1-\mathcal{F}(\bar{\epsilon}))}\int_{\bar{\epsilon}}^{\infty} \left(x - \bar{\epsilon}\right)d\mathcal{F}(x)\nonumber\\ 
		&> \frac{q}{(8/9)(1-1/2)}\int_{\frac{1}{2}}^{\infty} \left(x - \frac{1}{2}\right)d\mathcal{F}(x) = 2.25q\int_{\frac{1}{2}}^{\infty} \left(x - \frac{1}{2}\right)d\mathcal{F}(x),\label{eq:IA_caseIIa_for_SD_vs_IA}
	\end{align}
	where the inequality follows from $\bar{\epsilon} < \nicefrac{1}{2}$.
	
	Note that $\frac{2-3x}{(1-x)^2} \leq 2.25$ for $x \in [0,1/2]$. Thus, RHS of equation (\ref{eq:SD_for_IA_vs_SD}) is smaller than RHS of both equations (\ref{eq:IA_caseIb_for_SD_vs_IA}) and (\ref{eq:IA_caseIIa_for_SD_vs_IA}). As $c(x)$ is an increasing function, it means $\gamma^{IA} > \gamma^{Eq}$, a contradiction to my initial assumption.
\end{proof}

\begin{corollary}\label{corr:RD_IA}
	If the cost function $c(x)$ is such that 
	\begin{equation}
		c(0) < 2q \int_{\frac{1}{2}}^{\infty} \left(x - \frac{1}{2}\right) d\mathcal{F}(x) \label{eq:SD_vs_IA_utility1}
	\end{equation}
	and
	\begin{equation}
		c(2/3) \geq 3q\int_{0}^{\infty} x d\mathcal{F}(x), \label{eq:SD_vs_IA_utility2}
	\end{equation}
	then uninformed applicants have higher utility in the equilibrium of RSD than in the equilibrium of IA in which schools $A$ and $B$ are full after round 1.
\end{corollary}

\begin{proof}[Proof of Corollary \ref{corr:RD_IA}]
	Note that equation (\ref{eq:SD_vs_IA_utility1}) implies that some applicants are informed in RSD (see equation (\ref{eq:diffutil})) and equation (\ref{eq:SD_vs_IA_utility2}) implies that no more than $\nicefrac{2}{3}$ of applicants are informed in IA (see case IIa in Lemma \ref{lemma_IA-ABfull}).
	
	At $\gamma=0$, IA and RSD yield the same utility. As cutoffs $r(A;\gamma), r(B;\gamma)$ are increasing, for $\gamma > 0$ $U(0,ABC|r(A;\gamma),r(B;\gamma)) > U(0,ABC|r(A;0),r(B;0))$. When $\gamma \leq 2/3$, utility of an uninformed applicant under IA assignment is unchanged and equal to $U(0,ABC|r(A;0),r(B;0))$, as cutoffs $\rho_A$ and $\rho_B$ do not change when $\gamma \in [0,2/3]$. Thus, utility of an uninformed applicant is higher under RSD than under IA.
\end{proof}

\subsection{Non-symmetric distribution and three schools}\label{sec:appendix2}

	In this section, I revisit the conclusion of Section \ref{sec:SO} that information acquisition in the equilibrium of RSD game is below the socially optimal level for the case where $q_A \leq q_B$ and the distribution is not symmetric. Denote the probability $\epsilon_i > \nicefrac{1}{2}$ by $p_A$, the probability $\epsilon_i < -\nicefrac{1}{2}$ by $p_C$. Equations (\ref{eq:general-rA}), (\ref{eq:r_B_general_v}) and (\ref{eq:intuition1}) become
	\begin{align}
		&q_A = 
		r(A;\gamma) \left(1- \gamma p_A\right),\label{eq:very-general-rA}
		\\
		&q_B = 
		r(B;\gamma)\gamma p_A + (r(B;\gamma)-r(A;\gamma))\big(1-\gamma(p_A + p_C)\big).\label{eq:r_B_general_v2}
		\\
		&q_A + q_B = r(B;\gamma)(1 - \gamma p_C) + r(A;\gamma)\gamma p_C.
	\end{align}		
    While equation (\ref{eq:U(0|gamma)}) remains the same, (\ref{eq:U(1|gamma)-ver}) becomes 
	\begin{align}
		U&(1,\hat R(\cdot)|r(A;\gamma),r(B;\gamma)) = r(A;\gamma) \left(p_C + (1-p_A-p_C)E\left[\frac{1}{2}-\epsilon_i\Big|\epsilon_i\in \left[-\nicefrac{1}{2},\nicefrac{1}{2}\right]\right]\right)\nonumber
		\\&+ r(B;\gamma)\left(\frac{1}{2} + p_A E\left[\epsilon_i-\frac{1}{2}\Big|\epsilon_i\in \left[\nicefrac{1}{2},\infty\right]\right]\right)\nonumber
 		\\
		\begin{split}\label{eq:sorting-general}
		     =&\frac{q_A}{1-\gamma p_A} \left(p_C + (1-p_A-p_C)E\left[\frac{1}{2}-\epsilon_i\Big|\epsilon_i\in \left[-\nicefrac{1}{2},\nicefrac{1}{2}\right]\right]\right)
		     \\&+ \frac{q_A+q_B}{1-\gamma p_C}\left(\frac{1}{2} + p_A E\left[\epsilon_i-\frac{1}{2}\Big|\epsilon_i\in \left[\nicefrac{1}{2},\infty\right]\right]\right)
		\end{split}
		\\&- \frac{\gamma p_C}{1-\gamma p_C}\frac{q_A}{1-\gamma p_A}\left(\frac{1}{2} + p_A E\left[\epsilon_i-\frac{1}{2}\Big|\epsilon_i\in \left[\nicefrac{1}{2},\infty\right]\right]\right),\label{eq:displacement-general}
    \end{align}
    where the changes of (\ref{eq:sorting-general}) and (\ref{eq:displacement-general}) in $\gamma$ define sorting and displacement externalities.
	
	First, note that Theorem~\ref{th:Gamma_nonempty}, which establishes the existence of equilibrium, and Theorem~\ref{th:eqSD}, which describes the equilibrium of the RSD game, still apply, as they rely on the continuity of functions $r(A;x)$ and $r(B;x)$, but not on their particular form. 
    
	The derivative of $SW(\gamma^{Eq})$ (given by (\ref{eq:dSW}), which does not rely on symmetry) is 
	\begin{align*}
		\frac{\partial}{\partial \gamma} SW(\gamma^{Eq})
		= \gamma^{Eq} \frac{\partial}{\partial \gamma} U(1,\hat{R}(\cdot)|\gamma^{Eq}) + (1-\gamma^{Eq}) \frac{\partial}{\partial \gamma}U(0,ABC|\gamma^{Eq}).
	\end{align*}

	Algebraic manipulations show that $\frac{\partial}{\partial \gamma}r(A;\gamma) > 0$, $\frac{\partial}{\partial \gamma}(r(A;\gamma)+r(B;\gamma)) > 0$, with the latter inequality implying that a sorting externality dominates displacement for uninformed applicants, as in the symmetric model. Thus, the externalities on informed applicants determine whether information is under-acquired. Then, given that $\frac{\partial}{\partial \gamma}r(A;\gamma) > 0$, and in the absence of restrictions on $\mathcal{F}(x)$, the sufficient condition is given by the derivative of $r(B;\gamma)$:
	\begin{gather}
		\frac{\partial}{\partial \gamma}r(B;\gamma) = q_A p_C \frac{(1 - 2\gamma p_A)^2 + \gamma^2 p_A (p_C - 2 p_A))}{(1 - \gamma p_A)^2(1 - \gamma p_C)^2} + \frac{(q_B - q_A) p_C}{(1 - \gamma p_C)^2}
	\end{gather}
	Then \[\frac{\partial}{\partial \gamma}r(B;\gamma) > 0 \text{ when } p_C > \frac{1-2(1-p_A)^2}{p_A}.\] We can summarize the manipulations as the following claim:
	\begin{claim}\label{claim:app2_spec_dist}
		$\gamma^{SO} > \gamma^{Eq}$, if:
		\begin{enumerate}
			\item $p_C = 0$; \\or \label{claim:app2_spec_dist_case1}
			\item $p_C > \frac{1-2(1-p_A)^2}{p_A}$.\label{claim:app2_spec_dist_case2}
		\end{enumerate}
	\end{claim}
	Note that condition \ref{claim:app2_spec_dist_case2} implies that if $p_A < 1 - \sqrt{\nicefrac{1}{2}} \approx 0.3$, then $p_C$ can be arbitrary. Also note that there appear to be a discontinuity around $p_C = 0$: when $p_C$ is near zero, under-acquisition is less likely, but when $p_C = 0$, then under-acquisition is certain. In fact, there is no discontinuity. When $p_C$ is close to zero, the change in the cutoff $r(B;\gamma)$ is very small; however, with no restrictions on the distribution, that change can be magnified by increasing the tail of the distribution. When $p_C = 0$, the change in the cutoff is zero and unaffected by the choice of the distribution.  

\subsection{$N$-school environment}\label{subsec:appendix3}

In this section I study information acquisition in an environment with an arbitrary number of schools, but with a restriction on the distribution of applicant's realized utilities. I show that the main result reported in the paper (Theorem \ref{th:SO-more-than-DA}) extend to this environment.

\subsubsection{Environment}

A mass 1 of applicants is allocated to $N$ schools $s_1,\dots,s_i,\dots,s_N$ with quotas $q_1,\dots,q_i,\dots,q_N$ such that $\sum_{j=1}^{N-1}q_j < 1$ and $\sum_{j=1}^{N}q_j \geq 1$.

Applicant $i$'s utility for school $s_k$ is
\[
    u_i(s_k) = 
    \begin{cases}
        U(s_k) + \Delta^+_k &\text{ with probability } \pi\\
        U(s_k) - \Delta^-_k &\text{ with probability } (1-\pi),
    \end{cases}
\]
where 
\begin{enumerate}
    \item $\Delta^+_k > 0$, $\Delta^-_k>0$ and $\pi \times \Delta^+_k - (1-\pi) \times \Delta^-_k = 0$ for every $k \in \{1,\dots,N\}$; 
    \item $U(s_k) > U(s_{k+1})$  for every $k \in \{1,\dots,N-1\}$; and
    \item $U(s_k) + \Delta^+_k > U(s_{k-1}) - \Delta^-_{k-1} > U(s_{k+1}) + \Delta^+_{k+1}$ for every $k \in \{2,\dots,N-1\}$, $U(s_1) + \Delta^+_1 > U(s_{2}) + \Delta^+_{2}$ and $U(s_N) + \Delta^+_N > U(s_{N-1}) - \Delta^-_{N-1}$.
    \item $U(s_{k}) + \Delta^-_{k} > U(s_{k+1}) + \Delta^-_{k+1}$
\end{enumerate}

The first condition means that $U(s_k)$ is every applicant's expected utility from $s_k$. Second means that, in expectation, schools are decreasing in desirability from $s_1$ to $s_N$. Third means that utilities of only two adjacent schools can flip. That is, rankings $s_2s_1s_3$ or $s_1s_3s_2$ happen with probability \[\pi(1-\pi)\defeqi p,\] but $s_3s_1s_2$ or $s_3s_2s_1$ cannot happen. The last condition limits relative values of $\Delta^-_k$.

All these conditions are satisfied in a ``symmetric case'', where,  for any $k, j \in \{1,\dots,N\}$, $\pi = \nicefrac{1}{2}$, $U(k) - U(k+1) = U(j) - U(j+1)$ and $\Delta^+_k = \Delta^-_k = \Delta^+_{k+1} \in \left(\frac{U(k)-U(k+1)}{2},U(k)-U(k+1)\right)$.

Applicant $i$ knows expected utility $U(s_k)$ for every school, but does not know the realization; $i$ can learn the realizations for all schools at cost $c(i)$. I assume that $c(i)$ is strictly increasing.

I impose the following additional condition on school quotas:
\begin{align}
    \sum_{j=1}^k \left(-\frac{p}{1-p}\right)^{k-j}q_j > 0
\end{align}
for any $k \in \{1,\dots,N\}$. Note that, as $p<\nicefrac{1}{2}$, this condition is satisfied if quotas are non-decreasing in the expected desirability of schools: $q_{k} \geq q_{k-1}$ for any $k \in \{2,\dots,N\}$. This condition guarantees that school $s_k$ has a higher admission standard than school $s_{k+1}$ regardless of the fraction of informed applicants. In other words, it rules out the situation where the admission standards of two schools are inverted when applicants become more informed.\footnote{For an example of such environment, consider a two-school environment, where $U(s_1) > U(s_2)$, $q_1 = 0.9, q_2 = 0.1$ and $p=1/4$. When no one is informed, $r(s_1;0) = 0.9$ and $r(s_2;0) = 1$. When everyone is informed, 1/4 of applicants prefer $s_2$ to $s_1$; thus $r(s_2;1)=0.1/0.25=0.4$ and $r(s_1;1)=1$.}

\subsubsection{Main Result}

\begin{theorem}\label{th:multi-main}
    Suppose $\gamma^{Eq}< 1$. If
    \begin{align}\label{th:multi-condition}
        \gamma^{Eq} p < \min_{k \in \{2,\dots,N\}}\frac{q_k}{2q_{k-1} + q_k},
    \end{align}
    then $\gamma^{Eq} < \gamma^{SO}$.
\end{theorem}

Note that if $q_k = q_j$ for any $k, j \in \{1,\dots,N\}$, then the condition becomes the familiar $\gamma^{Eq} p < \nicefrac{1}{3}$. Recall also that $p \leq \nicefrac{1}{4}$. Thus, if the capacities of less desirable schools are not too small -- that is, if $q_k > \nicefrac{2}{3} \ q_{k-1}$ -- then the Condition \ref{th:multi-condition} necessarily holds.

To prove Theorem \ref{th:multi-main}, I derive school cutoffs in Subsection \ref{subsec:multi-cutoffs} and verify that cutoffs are increasing in $\gamma$ in Subsection \ref{subsec:multi_incr_cutoffs}. I then calculate expected utility of an uninformed applicant and the gain in information acquisition in Subsection \ref{subsec:multi-utils}. Combining these calculations, I show that social welfare increases in $\gamma$ at the equilibrium level of information acquisition $\gamma^{Eq}$ in Subsection \ref{subsec:multi-SW} under the condition (\ref{th:multi-condition}) of the theorem. This completes the proof.

\subsubsection{Cutoffs}\label{subsec:multi-cutoffs}

Claim \ref{claim:truthful} that uninformed applicants rank schools according to their expected utilities extends trivially to this environment. Note that if no applicants learn ($\gamma = 0$), then the cutoff for any school $s_k$, denoted $r(s_k;0)$, is lower (that is, $s_k$ is more selective) than the cutoff for school $s_{k+1}$, $r(s_k;0) < r(s_{k+1};0)$. I first derive cutoffs assuming that $\gamma \in [0,1]$ is such that $r(s_k;\gamma) < r(s_{k+1};\gamma)$ holds for any $k \in \{1,\dots,N\}$; then I show that this condition indeed holds for any $\gamma \in [0,1]$.

To calculate the cutoffs, I follow the steps in Section \ref{sec:DAA}. First, any applicant who prefers $s_2$ to $s_1$ will never be accepted to $s_1$; all other applicants apply to $s_1$ as their first choice, leading to the following supply-demand equation:
\begin{align*}
    (1 - \gamma p)r(s_1;\gamma) = q_1,
\end{align*}
which I re-write in the form identical to a supply-demand equation for arbitrary school $s_k$ derived later, as follows:
\begin{align}\label{eq:multi-SuppDemd-1}
    (1 - 2\gamma p)r(s_1;\gamma) + \gamma p r(s_1;\gamma) = q_1.
\end{align}
Applicants to $s_2$ are either 
\begin{itemize}
    \item applicants rejected from $s_1$ who rank $s_2$ above $s_3$; their scores are distributed uniformly on $(r(s_1;\gamma),1]$, and 
    \item applicants whose first choice is $s_2$; their scores are distributed uniformly on $[0,1]$. 
\end{itemize}
Then, the supply-demand equation for school $s_2$ is
\begin{align}\label{eq:multi-SuppDemd-2}
    (1 - 2\gamma p)\big(r(s_2;\gamma)-r(s_1;\gamma)\big) + \gamma p r(s_2;\gamma) = q_2.
\end{align}
For an arbitrary school $s_k$, with $k \geq 3$, there are two groups of applicants: rejected from $s_{k-1}$ and rejected from $s_{k-2}$; by construction, there are no other applicants to $s_k$. Thus, the supply-demand equation is
\begin{align}\label{eq:multi-SuppDemd-3}
    (1 - 2\gamma p)\big(r(s_k;\gamma) - r(s_{k-1};\gamma)\big) + \gamma p \big(r(s_k;\gamma) - r(s_{k-2};\gamma)\big) = q_k
\end{align}
Summing equations (\ref{eq:multi-SuppDemd-1})--(\ref{eq:multi-SuppDemd-3}), I obtain
\begin{align*}
    (1 - 2\gamma p) r(s_k;\gamma) + \gamma p \big(r(s_k;\gamma) + r(s_{k-1};\gamma) \big) = \sum_{j=1}^k q_j,
\end{align*}
leading to a recurrent formula for cutoffs:
\begin{align}
    r(s_1;\gamma) &= \frac{1}{1 - \gamma p}q_1 \nonumber\\
    r(s_k;\gamma) &= \frac{1}{1 - \gamma p}\left(\sum_{j=1}^k q_j - \gamma p r(s_{k-1};\gamma)\right),\label{eq:Nschool-r(s_k)}
\end{align}
from which I obtain
\begin{align}\label{eq:multi-cutoffs}
    r(s_k;\gamma) = \frac{1}{1 - \gamma p}\sum_{j=1}^k \left(\left(-\frac{\gamma p}{1 - \gamma p}\right)^{k-j}\sum_{l=1}^jq_{s^l}\right).
\end{align}

Note that
\begin{align}
    &r(s_k;\gamma) - r(s_{k-1};\gamma)=
    \\
    &\sum_{j=1}^k \left(\left(-\frac{\gamma p}{1 - \gamma p}\right)^{k-j}\sum_{l=1}^jq_{s^l}\right) - \sum_{j=1}^{k-1} \left(\left(-\frac{\gamma p}{1 - \gamma p}\right)^{k-j-1}\sum_{l=1}^jq_{s^l}\right)=
    \\
    &\sum_{j=1}^k \left(-\frac{\gamma p}{1 - \gamma p}\right)^{k-j} q_j > 0,
\end{align}
where the inequality ``$>0$'' follows for any $\gamma \in [0,1]$ from the assumptions on school quotas.

\subsubsection{Conditions for increasing cutoffs}\label{subsec:multi_incr_cutoffs}
Differentiating (\ref{eq:multi-cutoffs}), I obtain
\begin{align}\label{eq:multi-derivative}
    \frac{\partial}{\partial \gamma}r(s_k;\gamma) = \frac{p}{(1 - \gamma p)^2}\sum_{j=1}^k (k-j+1)\left(-\frac{\gamma p}{1 - \gamma p}\right)^{k-j}q_{j}
\end{align}
Note that $\frac{\partial}{\partial \gamma}r(s_1;\gamma)>0$ for all $\gamma, p$ and $q_1$.

For $k \geq 2$, consider $j$ so that $(k-j)$ is even or zero. The sum of two terms corresponding to $j$ and $j-1$ is
\begin{gather*}
    (k-j+1)\left(-\frac{\gamma p}{1 - \gamma p}\right)^{k-j}q_{j} + (k-(j-1)+1)\left(-\frac{\gamma p}{1 - \gamma p}\right)^{k-(j-1)} q_{j-1}=
    \\
    \left(-\frac{\gamma p}{1 - \gamma p}\right)^{k-j} \frac{(k-j+1)q_j - \gamma p \left((k-j+1)q_j + (k-j+2) q_{j-1}\right)}{1 - \gamma p}
\end{gather*}
This sum is positive whenever 
\begin{align}\label{eq:multi-condition}
    \gamma p < \frac{(k-j+1)q_j}{(k-j+1)(q_{j-1} + q_j) + q_{j-1}}.
\end{align}

If $k$ is even, then for each even $j$, the sum of terms $j-1$ and $j$ is positive. All terms in (\ref{eq:multi-derivative}) are exhausted by $(j-1,j)$ pairs, implying  $ \frac{\partial}{\partial \gamma}r(s_2;\gamma) > 0$.

If $k$ is odd, then for each odd $j \geq 3$, the sum of terms $(j-1,j)$ is positive. The term corresponding to $j=1$, $\frac{p}{(1-\gamma p)^2} k \left(-\frac{\gamma p}{1 - \gamma p}\right)^{k-1}q_{1} > 0$ because $(k-1)$ is even.

Thus, if inequality (\ref{eq:multi-condition}) holds for every $k \in \{2,\dots,N\}$ and every $j \leq k$, then $\frac{\partial}{\partial \gamma}r(s_k;\gamma) > 0$.

Inequality (\ref{eq:multi-condition}) can be simplified to inequality (\ref{th:multi-condition}) in the statement of the theorem by observing that, for a fixed $j$, if inequality (\ref{eq:multi-condition}) holds for $k=j$, then it holds for any $k>j$.

\subsubsection{Expected utilities of uninformed and informed applicants}\label{subsec:multi-utils}

Denote the ROL of uninformed applicant $R^P=(s_1,\dots,s_N)$. The expected utility of an uninformed applicant is
\begin{align*}
    U(0,R^P|\gamma) &= U(s_1)r(s_1;\gamma) + U(s_2)\big(r(s_2;\gamma)-r(s_1;\gamma)\big) + \dots + U(s_N)\big(1-r(s_{N-1};\gamma)\big)\nonumber
    \\
    &= U(s_N) + \sum_{k=1}^{N-1} \big(U(s_{k})-U(s_{k+1})\big)r(s_k;\gamma)
\end{align*}
Similarly, I calculate the expected utility of informed applicant by identifying the gain from learning for applicant's score in each band $(r(s_{k-1};\gamma),r(s_k;\gamma)]$. For convenience, I set $r(s_0;\gamma)=0$.
\begin{align}
    U(1,\hat{R}|\gamma)& 
    \begin{aligned}[t]
    = \sum_{k=1}^{N-1}\bigg(U(0,R^P|\gamma) + p\big(U(s_{k+1})+\Delta^+_{k+1} - &(U(s_{k})-\Delta^-_{k})\big)\bigg)\\
        &\times \big(r(s_{k};\gamma)-r(s_{k-1};\gamma)\big)\\
    \end{aligned}\nonumber\\
    &+ U(s_N)\big(1-r(s_{N-1};\gamma)\big)\nonumber
    \\
    &= U(0,R^P|\gamma) + p \sum_{k=1}^{N-1}\bigg(U(s_{k+1})+\Delta^+_{k+1} - (U(s_{k}) - \Delta^-_{k})\bigg) \big(r(s_{k};\gamma)-r(s_{k-1};\gamma)\big)\nonumber
    \\
    &= U(0,R^P|\gamma) + 
    \begin{aligned}[t]
        p \sum_{k=1}^{N-2}\bigg(&\big(U(s_{k+1})+\Delta^+_{k+1} - (U(s_{k}) - \Delta^-_{k})\big)
        \\
        &-\big(U(s_{k+2})+\Delta^+_{k+2} - (U(s_{k+1}) - \Delta^-_{k+1})\big)\bigg) r(s_{k};\gamma)
    \end{aligned}\label{eq:multi-util-diff-term}
    \\
    &+p \big(U(s_N)+\Delta^+_N - (U(s_{N-1}) - \Delta^-_{N-1})\big) r(s_{N-1};\gamma) \nonumber
    \\
    &\coloneqq U(0,R^P|\gamma) + \Delta U \nonumber
\end{align}

By plugging in (\ref{eq:Nschool-r(s_k)}) into (\ref{eq:multi-util-diff-term}), we obtain a definition of externalities similar to (\ref{eq:U(1|gamma)-ver}), where the sorting is represented by the first term in (\ref{eq:Nschool-r(s_k)}), and the displacement by the second.

As in the main model, the fraction of applicants who learn is defined as
\[
    \gamma^{Eq} = 
    \begin{cases}
        c^{-1}(\Delta U) & \text{ if } \Delta U \in [c(0),c(1)]\\
        0 & \text{ if } \Delta U < c(0)\\
        1 & \text{ if } \Delta U > c(1)
    \end{cases}
\]

\subsubsection{Social Welfare}\label{subsec:multi-SW}

Social welfare is defined as
\begin{align*}
    SW(\gamma) = U(0,R^P|\gamma) + \gamma \Delta U(\gamma) - \int_0^\gamma c(i) di.
\end{align*}
Differentiating, I get
\begin{align*}
    \frac{\partial}{\partial \gamma} SW(\gamma) = \big[\Delta U(\gamma) - c(\gamma)\big] + \gamma \frac{\partial}{\partial \gamma} \Delta U(\gamma) + \frac{\partial}{\partial \gamma} U(0,R^P|\gamma).
\end{align*}
The term $\Delta U(\gamma^{Eq}) - c(\gamma^{Eq}) = 0$, so the sign of $\frac{\partial}{\partial \gamma} SW(\gamma^{Eq})$ is determined by
\begin{align}
    &\gamma^{Eq} \frac{\partial}{\partial \gamma} \Delta U(\gamma^{Eq}) + \frac{\partial}{\partial \gamma}U(0,R^P|\gamma^{Eq})\nonumber
    \\
    &=\sum_{k=1}^{N-2}\label{eq:multi-long_term}
    \begin{aligned}[t]
        \Bigg[&\big(U(s_{k})-U(s_{k+1})\big)
        \\
        &+ \gamma^{Eq} p 
        \begin{aligned}[t]
            \bigg(&\big(U(s_{k+1})+\Delta^+_{k+1} - (U(s_{k}) - \Delta^-_{k})\big)
            \\
            &-\big(U(s_{k+2})+\Delta^+_{k+2} - (U(s_{k+1}) - \Delta^-_{k+1})\big)\bigg)\Bigg]
        \end{aligned}
        \\
        &\times \frac{\partial}{\partial \gamma} r(s_{k};\gamma^{Eq})
    \end{aligned}    
    \\
    &+ \bigg(\big(U(s_{N-1}) - U(s_N)\big ) + \gamma^{Eq} p \big(U(s_N)+\Delta^+_N - (U(s_{N-1}) - \Delta^-_{N-1})\big)\bigg) r(s_{N-1};\gamma)\label{eq:multi-short_term}
\end{align}
Consider an individual $k$-th term in square brackets in (\ref{eq:multi-long_term}). If the term multiplied by $\gamma^{Eq}p$ \begin{equation}\label{eq:multi-individDeltaterm}
\bigg(\big(U(s_{k+1})+\Delta^+_{k+1} - (U(s_{k}) - \Delta^-_{k})\big) - \big(U(s_{k+2})+\Delta^+_{k+2} - (U(s_{k+1}) - \Delta^-_{k+1})\big)\bigg)
\end{equation}
is positive, then the whole term in the square brackets is positive. Suppose then that (\ref{eq:multi-individDeltaterm}) is negative and consider the whole $k$-th term; recall that $\gamma \leq 1$ and $p < 1/2$.
\begin{align*}
    &\big(U(s_{k})-U(s_{k+1})\big)
    + \gamma^{Eq} p 
    \begin{aligned}[t]
        \bigg(&\big(U(s_{k+1})+\Delta^+_{k+1} - (U(s_{k}) - \Delta^-_{k})\big)
        \\
        &-\big(U(s_{k+2})+\Delta^+_{k+2} - (U(s_{k+1}) - \Delta^-_{k+1})\big)\bigg)
    \end{aligned}
    \\
    &> 
    \big(U(s_{k})-U(s_{k+1})\big)
    + \frac{1}{2} 
    \begin{aligned}[t]
        \bigg(&\big(U(s_{k+1})+\Delta^+_{k+1} - (U(s_{k}) - \Delta^-_{k})\big)
        \\
        &-\big(U(s_{k+2})+\Delta^+_{k+2} - (U(s_{k+1}) - \Delta^-_{k+1})\big)\bigg)
    \end{aligned}
    \\
    &= 
    \frac{1}{2}\bigg(U(s_{k}) + \Delta^-_k - (U(s_{k+1}) + \Delta^-_{k+1}) + U(s_{k+1}) + \Delta^+_{k+1}
    -(U(s_{k+2})+\Delta^+_{k+2})\bigg)
    > 0,
\end{align*}
where the last inequality follows from the assumptions on applicants' utilities. Thus, the sign of the term (\ref{eq:multi-long_term}) is the same as the sign of $\frac{\partial}{\partial\gamma} r(s_k;\gamma^{Eq})$. Similarly, term (\ref{eq:multi-short_term}) is larger than
\[
    \frac{1}{2}\bigg(U(s_{N-1}) - U(s_N) + \Delta^-_{N-1} + \Delta^+_N\bigg) > 0.
\]
Thus, $\frac{\partial}{\partial \gamma} SW(\gamma^{Eq}) > 0$ if $r(s_k;\gamma^{Eq}) > 0$ for all $k = \{1,\dots,N\}$; the latter holds when (\ref{th:multi-condition}) holds.

\subsection{Additional simulation results}\label{sec:appendix4}

An environment where a school has a very high variance -- that is, can be ranked as a top school by many students but as a bottom school by many others -- is less realistic than an environment without such a large swing. In this section, I provide the answer to the following question: ``What are the low-variance environments where we observe $\gamma^{SO} \leq \gamma^{Eq}$?'' Specifically, for each $N$ and each distribution, normal or uniform, I select environments -- determined by the vector of variances for each school, $(\sigma_1,\dots,\sigma_N)$ -- with $\gamma^{SO} \leq \gamma^{Eq}$. In Table \ref{tab:lowvar}, I report the environments where the $\max\{\sigma_1,\dots,\sigma_N\}$ is the lowest. 
A notable observation is that the environments where second-ranked school has high variance and other schools are certain -- similar to my three-school model -- are present in the table for each $N$. 

The intuition for this observation is as follows. If there are many schools with high variances, then ordinal preferences of informed applicants are diverse and applicants' sorting is a dominant consideration. When there is a high-variance school among low-variance schools, the requirement that the ordering of schools by selectivity does not change --  that is, for any $k < N$ and any $\gamma \in [0,1]$, $r(s_k;\gamma) < r(s_{k+1};\gamma)$ -- is often violated. The tuple $(0,0,5,\dots)$ for normal distribution is one example where $s_3$ is about to become more selective than $s_2$. There is only a narrow range of $\gamma^{Eq}$ -- between 0.86 and 0.865 -- where over-acquisition occurs, yet still $r(s_2;\gamma^{Eq}) < r(s_3;\gamma^{Eq})$. While the ordering between $s_1$ and $s_2$ never changes when distribution is symmetric and capacities are identical, the ordering between $s_2$ and $s_3$ may change because, when $\sigma_{3}$ is high but $\sigma_{1}$ and $\sigma_{2}$ are low, many applicants rank $s_3$ above \emph{both} $s_1$ and $s_2$. While demand for $s_1$ is always higher than for $s_3$, the \emph{residual} demand for $s_2$ may well be below the demand for $s_3$. Thus, the case where $s_2$ is the only high variance school is prominent in Table~\ref{tab:lowvar} because the condition that requires no change in the selectivity ordering does not restrict $\sigma_2$.

If one considers an environment with a mix of high- and low-variance schools unrealistic, the message from simulations is that the environments which do not violate school selectivity assumption are likely to have under-acquisition of information.

\begin{table}[h!]
	\centering \footnotesize
	\caption{Environments with lowest standard deviations and $\gamma^{SO} < \gamma^{Eq}$}\label{tab:lowvar}
	\begin{tabular}{ccccccccccccccc}
		\toprule
		\multicolumn{7}{c}{Normal distributions $\mathcal{F}_{s_k}(x)$} && \multicolumn{7}{c}{Uniform distributions $\mathcal{F}_{s_k}(x)$}
\\
\cline{1-7}\cline{9-15}
\\
\cline{2-7}\cline{10-15}
			{$\gamma^{Eq}$ in}& $\sigma_1$ & $\sigma_2$ & $\sigma_3$ & $\sigma_4$ & $\sigma_5$ & $\sigma_6$ & &  {$\gamma^{Eq}$ in}& $\sigma_1$ & $\sigma_2$ & $\sigma_3$ & $\sigma_4$ & $\sigma_5$ & $\sigma_6$ 
\\
 $[0.985,1]$ & 0 & 5 & 0 & -- & -- & -- & & $[0.985,1]$ & 0 & 4 & 0 & -- & -- & -- \\
 $[0.985,1]$& 0 & 5 & 1 & -- & -- & -- & &$[0.98,1]$& 0 & 4 & 1 & -- & -- & -- \\
 $[0.995,1]$& 1 & 5 & 0 & -- & -- & -- & &$[0.995,1]$& 1 & 4 & 0 & -- & -- & -- \\
	\cline{1-7}\cline{9-15}
 $[0.86,0.865]$& 0 & 0 & 5 & 0 & -- & -- & &$[0.985,1]$& 0 & 4 & 0 & 0 & -- & -- \\
 0.86  & 0 & 0 & 5 & 1 & -- & -- & &$[0.995,1]$& 1 & 4 & 0 & 0 & -- & -- \\
	\cline{9-15}
 $[0.985,1]$& 0 & 5 & 0 & 0 & -- & -- & &$[0.985,1]$& 0 & 4 & 0 & 0 & 0 & -- \\
 1 & 0 & 5 & 0 & 1 & -- & -- & &$[0.985,1]$& 0 & 4 & 0 & 3 & 0 & -- \\
 $[0.995,1]$& 1 & 5 & 0 & 0 & -- & -- & &$[0.975,1]$& 0 & 4 & 0 & 3 & 1 & -- \\
	\cline{1-7}
 $[0.86,0.865]$& 0 & 0 & 5 & 0 & 0 & -- & &$[0.98,1]$& 0 & 4 & 1 & 3 & 0 & -- \\
 $[0.985,1]$& 0 & 5 & 0 & 0 & 0 & -- & &$[0.995,1]$& 1 & 4 & 0 & 0 & 0 & -- \\
 $[0.995,1]$& 1 & 5 & 0 & 0 & 0 & -- & &$[0.995,1]$& 1 & 4 & 0 & 3 & 0 & -- \\
	\cline{1-7}
 $[0.86,0.865]$& 0 & 0 & 5 & 0 & 0 & 0 & &$[0.99,1]$& 1 & 4 & 0 & 3 & 1 & -- \\
	\cline{9-15}
 $[0.985,1]$& 0 & 5 & 0 & 0 & 0 & 0 & &$[0.985,1]$& 0 & 4 & 0 & 0 & 0 & 0 \\
 $[0.995,1]$& 1 & 5 & 0 & 0 & 0 & 0 & &$[0.985,1]$& 0 & 4 & 0 & 3 & 0 & 0 \\
  & & & & & & & &$[0.995,1]$& 0 & 4 & 0 & 3 & 0 & 1 \\
  & & & & & & & &$[0.995,1]$& 0 & 4 & 0 & 3 & 1 & 0 \\
  & & & & & & & &$[0.995,1]$& 0 & 4 & 0 & 3 & 1 & 1 \\
  & & & & & & & &$[0.98,1]$& 0 & 4 & 1 & 3 & 0 & 0 \\
  & & & & & & & &$[0.995,1]$& 0 & 4 & 1 & 3 & 0 & 1 \\
  & & & & & & & &$[0.99,1]$& 1 & 4 & 0 & 0 & 0 & 0 \\
  & & & & & & & &$[0.995,1]$& 1 & 4 & 0 & 3 & 0 & 0 \\
	\bottomrule
	\end{tabular}%
	
	\begin{tabnotes}
		For each environment with $N$ school and either normal or uniform distribution, I find the tuple $(\sigma_{1}, \dots, \sigma_{N})$ where (i) $\gamma^{SO} < \gamma^{Eq}$ and (ii) the largest $\sigma_{k}$ is the lowest (which is 5 for normal and 4 for uniform distributions across all $N$-school environment, even though each is treated independently). I then report the interval of $\gamma^{Eq}$  and all tuples that satisfy these conditions. $\gamma^{Eq}$ refers to the equilibrium fraction of informed applicants; $\gamma^{SO}$ is to the social optimum given $c(i) = c \times i$ corresponding to $\gamma^{Eq}$.
	\end{tabnotes}	
\end{table}%

\ifx\undefined\BySame
\newcommand{\BySame}{\leavevmode\rule[.5ex]{3em}{.5pt}\ }
\fi
\ifx\undefined\textsc
\newcommand{\textsc}[1]{{\sc #1}}
\newcommand{\emph}[1]{{\em #1\/}}
\let\tmpsmall\small
\renewcommand{\small}{\tmpsmall\sc}
\fi

\end{document}